\documentclass[12pt]{JHEP3}

\def\hpolePercentage{35}
\def\stauCoanPercentage{4}
\def\fpPercentage{61}
\def\nonFPPercentage{39}

\title{Panglossian Prospects for Detecting Neutralino Dark Matter in Light of
Natural Priors?}

\author{Benjamin C.\ Allanach$^{1}$ and Dan Hooper$^{2,3}$ \\ 
$^{1}$ DAMTP, CMS, University of Cambridge, Wilberforce Road, Cambridge, CB3
  0WA, UK\\ 
$^{2}$ Fermi National Accelerator Laboratory, Theoretical Astrophysics, Batavia, IL 60510, USA\\
$^{3}$ The University of Chicago, Department of Astronomy and Astrophysics, Chicago, IL 60637, USA\\
}
\keywords{Supersymmetry, Effective Theories,
Cosmology of Theories Beyond the Standard Model,
Dark Matter}

\abstract{In most global fits of the
  constrained minimal supersymmetric model (CMSSM) to indirect data, the {\it
    a priori}\/ likelihoods of   any two points in $\tan \beta$ are treated as
  equal, and   the more fundamental $\mu$ and $B$ Higgs 
  potential parameters are fixed by potential minimization
  conditions. 
  We find that, if instead a flat (``natural'') 
  prior measure on $\mu$ and $B$ is placed, 
  a strong preference exists for the focus point region from 
  fits to particle physics and cosmological data. In particular, we
  find that the lightest neutralino is strongly favored to be a mixed
  bino-higgsino ($\sim 10\%$ higgsino). Such mixed neutralinos have large
  elastic scattering cross sections with nuclei, leading to extremely
  promising prospects for both underground direct detection experiments and
  neutrino telescopes. In particular, the majority of the posterior
  probability distribution falls within parameter space within an order of
  magnitude of current direct detection constraints. Furthermore, neutralino
  annihilations in the sun are predicted to generate thousands of neutrino
  induced muon events per years at IceCube. 
  Thus, assuming the framework of the CMSSM and using the natural
  prior measure, modulo caveats regarding astrophysical uncertainties,
  we are likely to be living in a
  world with good prospects for the direct and indirect detection of
  neutralino dark matter.
} 

 \newlength{\wth}
\usepackage[dvips]{epsfig}

\newcommand{\twographs}[2]{%
 \unitlength=1.1in
 \begin{picture}(5.8,2.3)(0.5,0.25)
 \put(-0.04,2.54){\epsfig{file=#1, width=0.698 \wth,angle=270}}
 \put(0.85,0.5){\epsfig{file=#12, width=0.68 \wth}}
 \put(2.66,2.54){\epsfig{file=#2, width=0.698 \wth, angle=270}}
 \put(3.56,0.5){\epsfig{file=#22, width=0.68 \wth}}
 \end{picture}
}

\newcommand{\twographsb}[2]{%
 \unitlength=1.1in
 \begin{picture}(5.8,2.3)(0.5,0.25)
 \put(-0.04,2.54){\epsfig{file=#1, width=0.698 \wth,angle=270}}
 \put(0.52,0.5){\epsfig{file=#12, width=0.68 \wth}}
 \put(2.66,2.54){\epsfig{file=#2, width=0.698 \wth, angle=270}}
 \put(3.26,0.5){\epsfig{file=#22, width=0.68 \wth}}
 \end{picture}
}

\newcommand{\twographsCDMS}[2]{%
 \unitlength=1.1in
 \begin{picture}(5.8,2.3)(0.5,0.25)
 \put(-0.04,2.54){\epsfig{file=#1, width=0.698 \wth,angle=270}}
 \put(0.85,0.5){\epsfig{file=#12, width=0.68 \wth}}
 \put(0.99,1.58){\epsfig{file=#13, width=0.42 \wth}}
 \put(2.66,2.54){\epsfig{file=#2, width=0.698 \wth, angle=270}}
 \put(3.56,0.5){\epsfig{file=#22, width=0.68 \wth}}
 \put(3.70,1.58){\epsfig{file=#23, width=0.42 \wth}}
 \end{picture}
}

\newcommand{\twographsGLAST}[2]{%
 \unitlength=1.1in
 \begin{picture}(5.8,2.3)(0.5,0.25)
 \put(-0.04,2.54){\epsfig{file=#1, width=0.698 \wth,angle=270}}
 \put(0.85,0.5){\epsfig{file=#12, width=0.68 \wth}}
 \put(0.97,1.60){\epsfig{file=#13, width=0.43 \wth}}
 \put(2.66,2.54){\epsfig{file=#2, width=0.698 \wth, angle=270}}
 \put(3.56,0.5){\epsfig{file=#22, width=0.68 \wth}}
 \put(3.68,1.60){\epsfig{file=#23, width=0.43 \wth}}
 \end{picture}
}

\preprint{DAMTP-2008-51\\ FERMILAB-PUB-08-153-A} 

\begin{document}

\section{Introduction}

For a variety of reasons, supersymmetry is considered to be among the most
attractive extensions of the Standard Model. In particular, weak-scale
supersymmetry provides an elegant solution to the hierarchy
problem~\cite{susyreview}, and enables grand unification by causing the gauge
couplings of the Standard Model to evolve to a common scale~\cite{gut}. From
the standpoint of providing a dark matter candidate, the lightest neutralino
is naturally stable by virtue of R-parity conservation \cite{neutralinodm},
and in many models is thermally produced in the early universe in a quantity similar to the measured density of cold dark matter~\cite{wmap5}. 

In addition to collider searches for superpartners, a wide range of
astrophysical experiments are currently operating and being developed in the
hopes of detecting neutralino dark matter~\cite{Bertone:2004pz}. These
techniques can be classified as direct and indirect detection. While the
former efforts are designed to observe the elastic scattering of neutralinos
with target nuclei, the latter techniques attempt to detect the annihilation
products of neutralinos, including gamma-rays~\cite{gamma},
neutrinos~\cite{neutrinos}, positrons~\cite{positron},
antiprotons~\cite{antiproton}, antideuterons~\cite{antideu}, and synchrotron
radiation~\cite{syn}. In addition to astrophysical inputs, the prospects for
direct and indirect dark matter detection depend on the mass and couplings of the lightest neutralino, and in turn on the many parameters which define the masses and couplings of the superpartners.

Weak-scale supersymmetry could take a great variety of forms, depending on the
details of how supersymmetry is broken. Empirically, our insights into this
question are limited to the measurements of observables indirectly related to
the supersymmetric spectrum, such as the anomalous magnetic moment of the
muon, the $b \rightarrow s \gamma$ branching fraction, the $B_s \rightarrow
\mu^+  \mu^-$ branching fraction, the mass of the $W$ boson, the effective
leptonic mixing angle, Higgs boson and sparticle search constraints, and the
cosmological dark
matter abundance. Such observables have been used in the past to constrain the
properties of the CMSSM spectrum (see, for example,
Refs.~\cite{ellis,Profumo:2004at,Allanach:2005kz,Roszkowski:2006mi,weather}). 
Ultimately,  
this information can be used to determine the posterior probability
distribution over the parameter space of supersymmetry. In
Refs.~\cite{deAustri:2006pe,Roszkowski:2007fd,Roszkowski:2007va},
it was used to examine the prospects for dark matter detection. 

In this paper, we consider another input that can play a significant role in
determining the posterior distribution over supersymmetric parameters. In
particular, we consider the measure which is associated with
each point in parameter space and define a prior measure which is flat in
terms of fundamental CMSSM parameters. In
our analysis, we closely follow Ref.~\cite{weather}, but focus on the
phenomenology of neutralino dark matter in the regions of supersymmetric
parameter space favored by indirect constraints {\em and}\/ naturalness
considerations. 
When a natural prior measure (flat in more fundamental CMSSM parameters, rather
than in $\tan \beta$) is
included in the analysis of the parameter 
space of the constrained minimal supersymmetric standard model (CMSSM), we
find that the focus point region is highly preferred. In this region, the
lightest neutralino $\chi_1^0$ is a mixed bino-higgsino ($\sim 10\%$ higgsino
fraction) 
and, therefore, has relatively significant couplings to the Standard Model.  

The prospects for the direct and indirect detection of neutralino dark matter
in the favored regions are highly promising. In particular,
about \fpPercentage\% of the posterior probability distribution predicts a
neutralino-nucleon elastic scattering cross section of $\sigma_{\chi^0 N}
\approx 10^{-8}-10^{-7}$ pb, which is within one order of magnitude of the
current direct detection constraints. 
The remaining \hpolePercentage\% of the posterior
probability distribution corresponds to parameter space in which the lightest neutralino has somewhat smaller couplings (and direct detection rates) but still
annihilates efficiently in the early universe via the light Higgs resonance
($2 m_{\chi^0} \approx m_h$). The projected rates at neutrino telescopes are also
extremely promising, with most of the posterior probability distribution being
made up of models which predict thousands of events per year at a
kilometer-scale neutrino telescope such as IceCube. Current constraints from
Super-Kamiokande and Amanda/IceCube already exclude a sizable fraction of the otherwise favored
probability distribution. We also discuss the prospects for indirect detection
using gamma-rays and charged cosmic ray particles.  

\section{The Measure of CMSSM Parameter Space}

The CMSSM parameter space consists of the following supersymmetry (SUSY)
breaking parameters: the universal scalar mass $m_0$, the universal gaugino mass $m_{1/2}$, and the universal tri-linear scalar
coupling $A_0$. These parameters constrain the SUSY breaking terms in the CMSSM
potential at a high energy scale, which is usually taken to be $M_{GUT}$, the
scale at which the electroweak gauge couplings unify. 
In addition, $\tan \beta$ is often used to characterize the ratio of the two
Higgs doublet vacuum expectation values and is taken to be an input parameter. 
When performing global fits to the CMSSM, it is important to take into account
any smearing due to variations in important Standard Model input parameters,
which we denote collectively as $s$. 
One defines the likelihood, $p(D, M_Z^{emp} | m_0, m_{1/2}, A_0, 
\tan \beta, s, M_Z)$, by calculating the probability density of the parameter point
reproducing all current data, $D$. We have singled out the
empirically measured $Z^0$ boson pole mass, $M_Z^{emp}$, and the one predicted
in the CMSSM, $M_Z$, since they have a special r\^{o}le in what follows. 

Here in contrast, 
in order to make probabilistic inferences, we begin by defining
a measure in the parameter space of the CMSSM by following Ref.~\cite{weather}: 
\begin{eqnarray}
p({D})&=& \int d \mu\ d B \ d  A_0 \ d m_0 \ d m_{1/2} \ ds\left[ 
p(m_0, m_{1/2}, A_0, \mu, B, s)\right. \nonumber \\ 
&& \left. p(D, M_Z^{emp} | m_0, m_{1/2}, A_0, \mu, B, s)
\right], \label{initial} 
\end{eqnarray}
where $p(m_0, m_{1/2}, A_0, B, \mu, s)$ is
the joint prior probability distribution for CMSSM and Standard Model
parameters. 
In fact, $M_Z$ and $\tan \beta$ are related to
the more fundamental 
parameters by the MSSM Higgs potential minimization conditions~\cite{BPMZ}:
\begin{eqnarray}
  \mu^2 &=& \frac{{\bar m}_{H_1}^2 - {\bar m}_{H_2}^2 \tan^2 \beta}{\tan^2
  \beta -1} -   \frac{M_Z^2}{2} \label{musq} \\ 
  \mu B &=& \frac{\sin 2 \beta}{2} ( {\bar m}_{H_1}^2 + {\bar m}_{H_2}^2 + 2
  \mu^2 ).
  \label{muB} 
\end{eqnarray}
Eq.~\ref{musq} is applied at a renormalization scale equal to the geometric
mean of the two stop masses, $Q\sim \sqrt{m_{{\tilde t}_1}
    m_{{\tilde t}_2}}$, which cancels some
larger logarithms in higher order corrections and results in higher accuracy.
${\bar m}_{H_1}^2$ and ${\bar m}_{H_1}^2$ are obtained from the
universality boundary condition on scalar masses at $M_{GUT}$. They are 
run to $Q$ and corrected by
some tadpole loop corrections~\cite{softsusy}. Since 
$M_Z^{emp}$ and the other data, $D$, are independent,
\begin{eqnarray}
p(D, M_Z^{emp} | m_0, m_{1/2}, A_0, \tan \beta, s)&=&
p(D | m_0, m_{1/2}, A_0, \tan \beta, s) \times \nonumber \\
&&p(M_Z^{emp} | m_0, m_{1/2}, A_0, \tan \beta, s). \label{factorise}
\end{eqnarray}
Direct current data imply that the $Z^0$
boson mass is extremely well constrained,
$M_Z^{emp}=91.1876\pm0.0021$~\cite{pdg06} , and so we make the approximation:
\begin{equation}
p(M_Z^{emp} | m_0, m_{1/2}, 
A_0, \tan \beta, s) \approx \delta (M_Z - M_Z^{emp}). \label{MZapprox}
\end{equation}

In the present paper, $p(m_0, m_{1/2}, A_0, \mu, B, s)$ is defined to be a
constant, resulting in so-called ``flat'' priors in the named parameters.  
Probabilistic inferences may be made based upon the {\em posterior}\/
probability distribution, defined to be the product of likelihood and prior,
integrated over all parameters except for the ones we are interested in, using
the previously defined measure. 
In most previous Bayesian global fits to the
CMSSM~\cite{Allanach:2005kz,deAustri:2006pe,Roszkowski:2007fd,Roszkowski:2007va,darkSide,allanachnatural},  
(often flat) prior probability distributions were defined in terms of the
measure 
\begin{equation}
dM \equiv d \tan \beta\ d M_Z\ dm_0\ d m_{1/2}\ d A_0\ d s. \label{oldMeasure}
\end{equation}
We refer to $dM$ as the ``flat $\tan \beta$'' measure if it is used in
conjunction with a prior probability distribution that is flat in each of the
parameters named on the right-hand side of Eq.~\ref{oldMeasure}.
One must be aware that different measures for the parameters may be
chosen, and will affect the results if the power of the data is weak. For
example, in Ref.~\cite{darkSide}, $dM$ priors that were flat in $\ln m_0$ and
$\ln  
m_{1/2}$ 
were compared to those that are flat in $m_0$ and $m_{1/2}$. In
Ref.~\cite{allanachnatural}, a naturalness prior was 
introduced in terms of $dM$ that  
disfavors regions of parameter space for which large cancellations are
necessary in the Higgs potential~\cite{finetuning}. 
In Ref.~\cite{weather}, $p(m_0, m_{1/2}, A_0, \mu, B, s)$ from
Eq.~\ref{initial} was 
chosen to strongly disfavor hierarchies between the different parameters,
encoding the prejudice that they should be of the same order. In this study, we
drop the ``of the same order'' prejudice, which was deemed by
Ref.~\cite{Roszkowski:2007fd} to be going a step too far. 
By comparing the results found in studies using different prior measures, some
non-negligible dependence upon the prior measure chosen can be found,
indicating that determinations of the favored regions of the CMSSM parameter
space from current data are somewhat uncertain. If more data compatible with
the CMSSM is obtained in the future, it is expected that this unwanted
dependence on the choice of the prior measure will be reduced.  

Following Ref.~\cite{weather}\footnote{Note that in
  Ref.~\protect\cite{weather}, the prior factor in Eq.~\ref{initial2} 
was called the ``REWSB'' prior. Here, we refer to it as a ``natural prior''.},
substituting Eqs.~\ref{MZapprox} 
and~\ref{factorise} into  
Eq.~\ref{initial}, and calculating the Jacobian of $d \mu\ d B \rightarrow 
d \tan \beta \ d M_Z$ from Eqs.~\ref{musq} and~\ref{muB}, we arrive at a map between $dM$ and our desired measure:
\begin{eqnarray}
p({D})&=& \int d \tan \beta \ d  A_0 \ d m_0 \ d m_{1/2} \ ds\left[ 
p(m_0, m_{1/2}, A_0, \mu, B, s)\right. \nonumber \\ 
&& \left. p(D | m_0, m_{1/2}, A_0, \mu, B, s) M_Z 
\left| \frac{B}{\mu \tan \beta} \frac{\tan^2 \beta - 1}{\tan^2 \beta+1} \right|
\right]_{M_Z=M_Z^{emp}}, \label{initial2} 
\end{eqnarray}
where $\mu$ and $B$ are obtained from Eqs.~\ref{musq} and~\ref{muB}. 
For now, until more data are obtained, we are stuck with dependence upon the
priors and so attempts to make good guesses for reasonable prior distributions
are important. The prior measure defined in Eq.~\ref{initial2} is clearly
superior to $dM$ because it is phrased in terms of parameters that are more
fundamental to the model: namely, $\mu$ and $B$ rather than $\tan \beta$ and
$M_Z$. 
We shall compare and contrast the posterior samples
obtained from these different priors. Note that one can still argue whether
the prior measure should be flat in $\mu$ and $B$, or whether some other
measure (such as one flat in $\log B$ and $\log \mu$ for instance, see the
discussion in Ref.~\cite{weather}) is more
appropriate. If a flat prior in $\log B$, $\log \mu$ is taken, one can
multiply the integrand of Eq.~\ref{initial2} by a further factor of 
$1/(B \mu)$. 
Whichever choice is taken, we believe that the connection with
the fundamental parameters of the MSSM is clearer if one starts from a measure
$d \mu \ d B$, rather than $d \tan \beta d M_Z$. 
We refer to the prior measure defined in
Eq.~\ref{initial2} with 
constant $p(m_0, m_{1/2}, A_0, \mu, B, s)$, as the natural prior. 

\section{Electroweak Symmetry Breaking and CMSSM Parameter Space}

Our calculation of the likelihood closely follows the calculation found in
Ref.~\cite{weather}, with additional $b-$physics observables and updated
empirical values.  
The four important Standard Model (SM) inputs referred to in the previous
section collectively as 
$s$ are: the inverse fine structure constant evaluated in the $\overline{MS}$
scheme at $M_Z$, $1/\alpha^{\overline{MS}}(M_Z)=127.918 \pm
0.018$~\cite{pdg06},
the equivalent version of the strong coupling constant,
$\alpha_s^{\overline{MS}}(M_Z)=0.1176\pm 0.002$~\cite{pdg06}, the bottom quark
mass evaluated at its own mass,
$m_b(m_b)^{\overline{MS}}=4.20\pm0.07$ GeV~\cite{pdg06}, and the pole top quark
mass, $m_t=172.6\pm 1.8$ GeV \cite{cdf+dzero-mtop-07}. The muon decay constant
is very accurately determined, and its central value is used as a fixed input,
$G_\mu=1.16637 \times 10^{-5}$ GeV$^{-2}$, and is used to predict the $W$ boson
pole mass, $M_W$.

\TABULAR{|l | c c | c|}{\hline
Observable & Central value & Combined Uncertainty &  References \\\hline
$R_{BR(B_u \rightarrow \tau \nu)}$ & $1.259$ & $0.378$ & \cite{btaunu} \\
$\Delta_{o-}$ & $0.0375$ & $0.0289$ & \cite{gamb2} \\
$R_{\Delta_{m_s}}$ & $0.85$ & $0.12$ & \cite{btaunu,delms_sm} \\
${\delta a_{\mu}} \times 10^{10}$ & $29.5$ & $8.8$ & \cite{gm2SM}\\
$M_W$ & $80.398$ GeV & $27$ MeV & \cite{mw} \\
$\sin^2 \theta_w^l$ & $0.23149$ & $0.000173$ &  \cite{sinth,mw2}  \\
$BR(b \rightarrow s \gamma)\times 10^{4}$ & $3.55$ & $0.72$ &
\cite{gamb}\\
$\Omega_{DM} h^2$ & 0.1143 & 0.01 & \cite{wmap5}\\
 \hline
}{Indirect constraints used. For each quantity, an estimate of the theoretical
  error in our CMSSM prediction has been added to the empirical error in
  quadrature.  
\label{tab:observables}}
In table~\ref{tab:observables}, we show the updated values of the observables used
in our likelihood calculation,
along with the relevant references.  Here, $R_{BR(B_u \rightarrow \tau 
  \nu)}$ is the  
ratio of the experimental and SM predictions of the branching ratio of $B_u$
mesons decaying into a tau and a tau neutrino. 
The SM prediction of this quantity is rather uncertain 
because of two incompatible empirically derived values of $|V_{ub}|$:
$(3.68 \pm 0.14) \times 10^{-3}$ versus 
the value coming from inclusive semi-leptonic
decays, $(4.49 \pm 0.33) \times 10^{-3}$. We simply combine these two
measurements assuming independent Gaussian errors to give our SM prediction of
the branching ratio
$BR^{\mathrm{SM}}(B_u \rightarrow \tau \nu) = (112 \pm 25) \times 10^{-6}$.
$R_{\Delta_{m_s}}$ is the ratio of the experimental and the SM neutral $B_s$
meson mixing  amplitudes. $\Delta_{0-}$ is the isospin asymmetry in $B
\rightarrow K^* \gamma$ decays.

We have used {\tt SOFTSUSY2.0.17}~\cite{softsusy} to calculate the sparticle
and Higgs masses and couplings. 
Any point in the CMSSM parameter space contravening 95$\%$ confidence level
sparticle direct search limits is given zero
likelihood as described in Ref.~\cite{darkSide}. The SM inference of
the LEP2 Higgs search may be used to constrain the lightest CP-even Higgs
boson $h^0$ of the CMSSM, since other constraints force the model to be in the 
decoupling SM-like r\'{e}gime. Thus, likelihood penalties from LEP2 are
combined with a 3 GeV Gaussian smearing to model the uncertainty in the
{\tt SOFTSUSY2.0.17} theoretical prediction. 
The SUSY Les Houches Accord~\cite{slha} is used to transfer the 
spectral information to
{\tt micrOMEGAs2.1}~\cite{micromegas}, which calculates the relic
density of neutralino dark matter and its elastic scattering and annihilation cross sections, and {\tt SuperIso2.0}~\cite{superiso}, which
calculates the branching ratio of $b$ quarks into $s$ quarks and a photon
using one-loop MSSM corrections and NNLO SM QCD corrections. {\tt SuperIso2.0}
is also used to predict $\Delta_{o-}$. $M_W$ and $\sin^2 \theta_w^l$ are
predicted with the full 
two-loop MSSM effects included~\cite{Heinemeyer:2006px}. 
$R_{\Delta_{m_s}}$ and $R_{BR(B_u \rightarrow \tau \nu)}$ are computed using
the approximate one-loop expressions in
Refs.~\cite{Buras:2002vd,Isidori:2006pk} respectively. 

We performed a Markov Chain Monte Carlo bank sampling scans~\cite{bank} over
four SM inputs and the four continuous CMSSM parameters, choosing $\mu>0$. 
There is a statistical preference coming from the $(g-2)_\mu$
measurement~\cite{darkSide}. Several chains were run 
using different random numbers
for 200,000 steps each. For each chain, 5000 bank points were obtained
at random from previous 10$\times$50,000 step scouting Metropolis runs.
In particular, it was important to include points from the $h$-pole region,
the stau co-annihilation region and the focus point region in the bank (all
described below) as these good-fit regions were (in some cases) not simply
connected, a situation ideally suited to bank sampling.
Enough chains were generated in order that they satisfy the Gelman
and Rubin 
convergence criterion of
$\sqrt{\hat{R}}<1.05$~\cite{GelmanAndRubin}. $\sqrt{\hat{R}}$ 
provides an estimated upper bound on the decrease in standard deviation that
could be obtained in any of the eight input parameters by running the MCMC
chains for more steps.  
9 chains were
sufficient for natural priors, whereas 20 were sufficient for the flat prior
case.  
Our scan was performed over the parameter ranges:  60 GeV $<$
$m_{1/2}$ $<$ 2 TeV, 60 GeV $<$ $m_{0}$ $<$ 4 TeV, -4 TeV $<$ $A_0$ $<$ 4 TeV,
2 $<$ $\tan \beta$ $<$ 62. Bank sampling allows us to efficiently sample from
distributions which have well separated peaks, which is the case for the
natural posterior probability distribution. 
 
\FIGURE{\includegraphics[width=2.9in,angle=0]{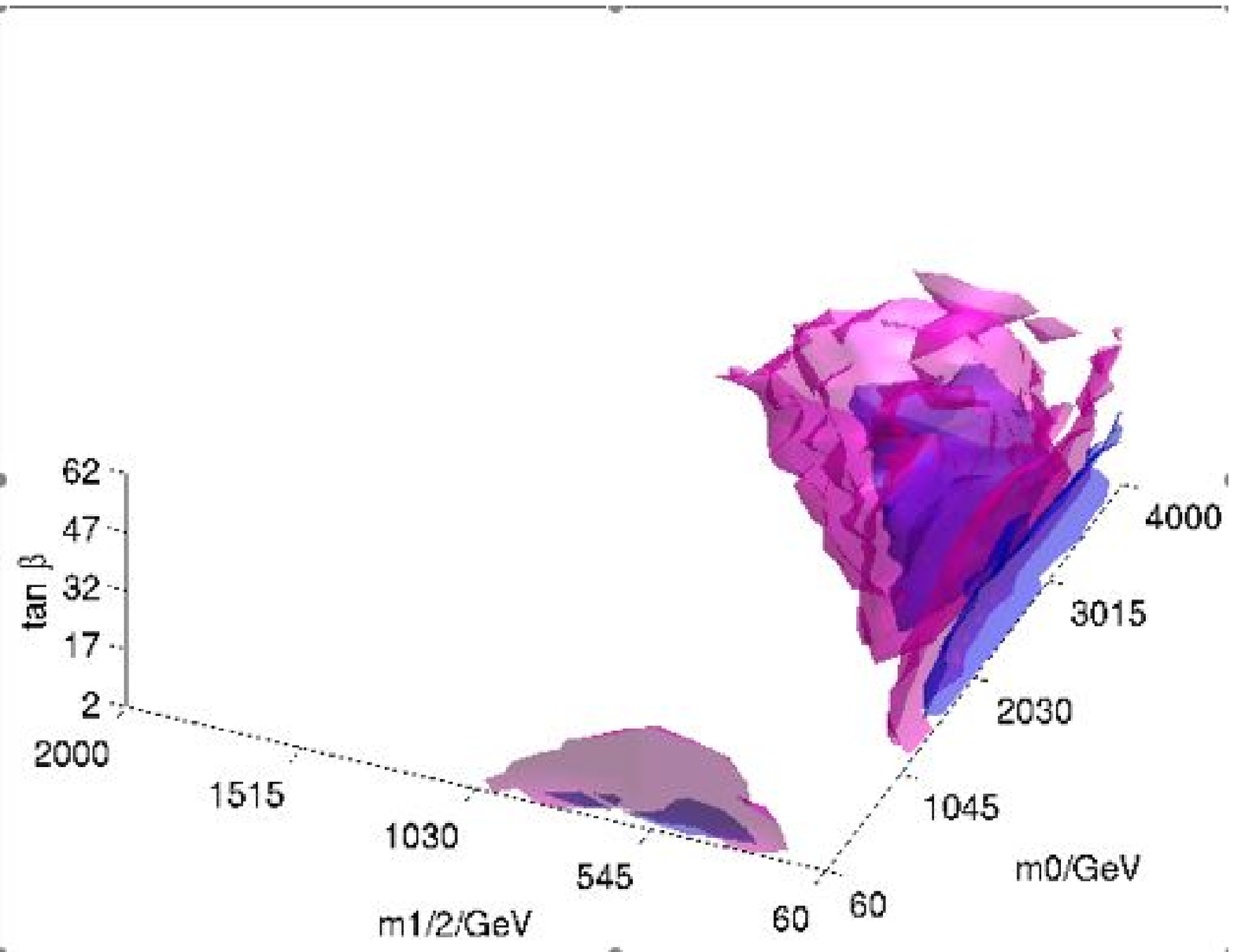}
\includegraphics[width=2.9in,angle=0]{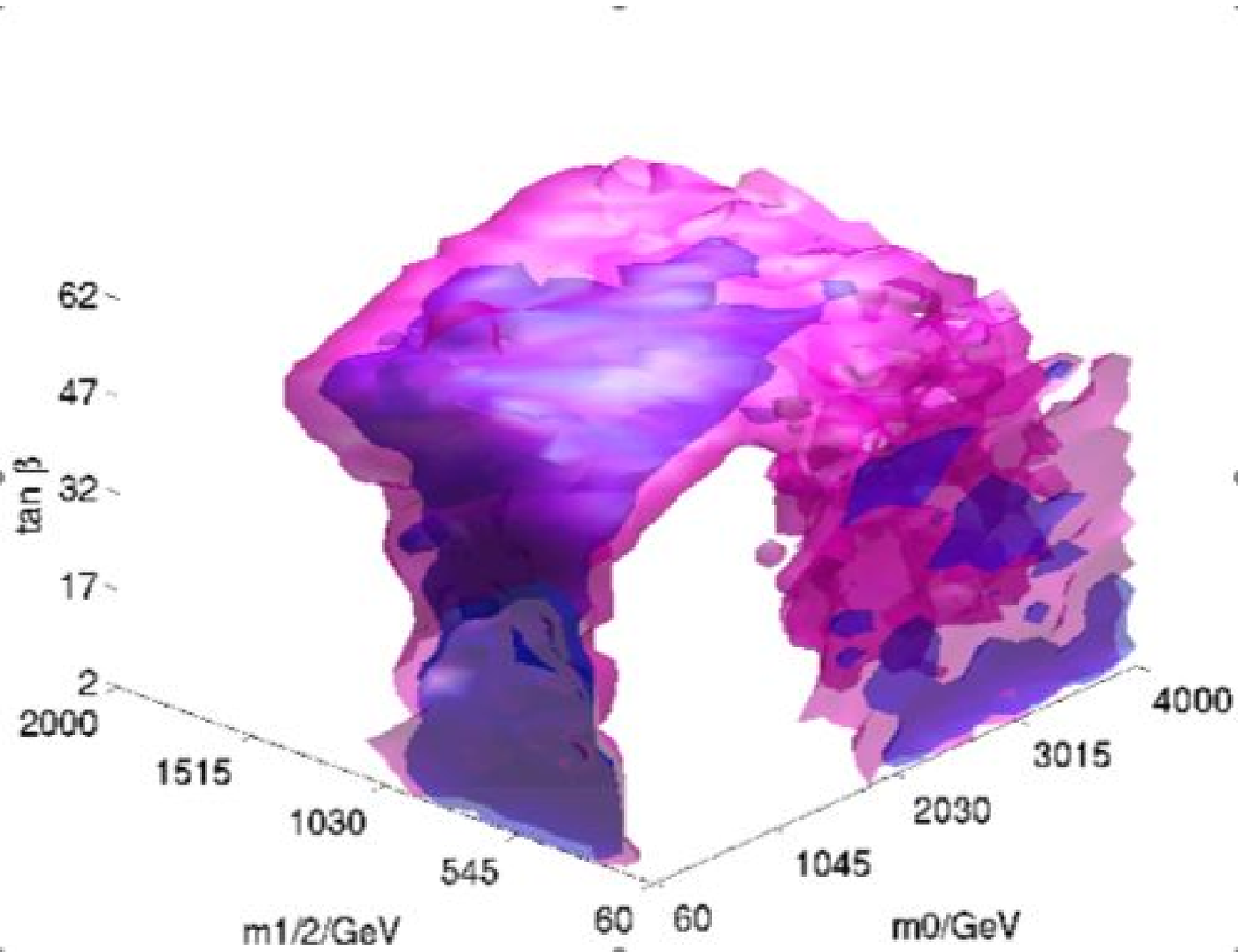}
\caption{Iso-posterior probability density surfaces of the CMSSM parameters,
  projected in three dimensions. The posterior has been marginalized over the
  unseen parameters, 
  taking into account the empirical inputs described in the text and using a
  natural prior (left) or the
  flat $\tan \beta$ prior (right) as
  described in the text. The inner (outer) surfaces contain 68$\%$(95$\%$)
  of the posterior probability density, respectively. The natural prior
  enhances the focus point region (bottom) for the reasons discussed in the
  text.} 
\label{3D}}

\FIGURE{\twographsb{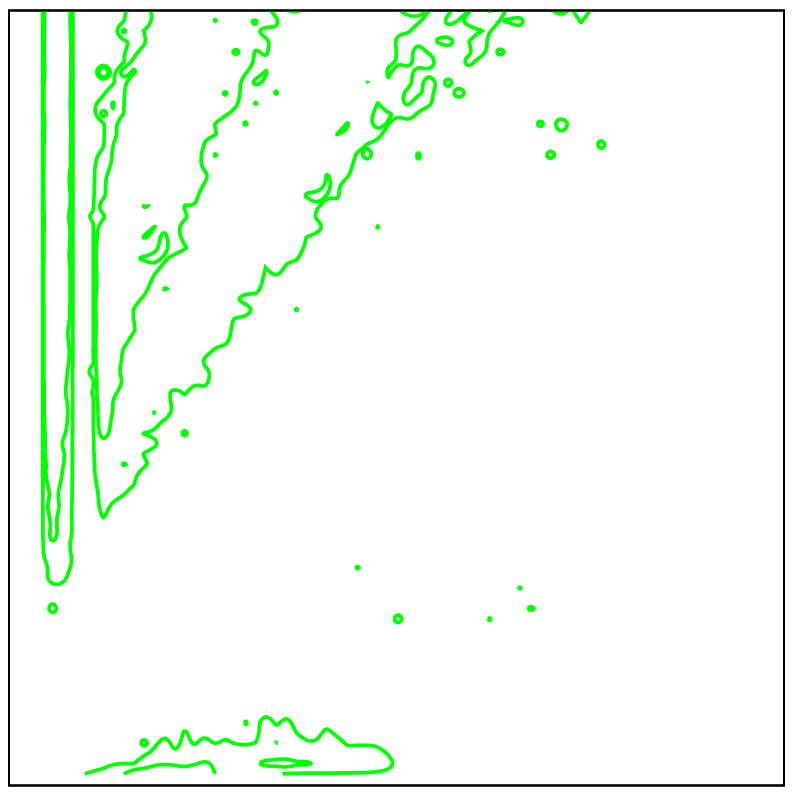}{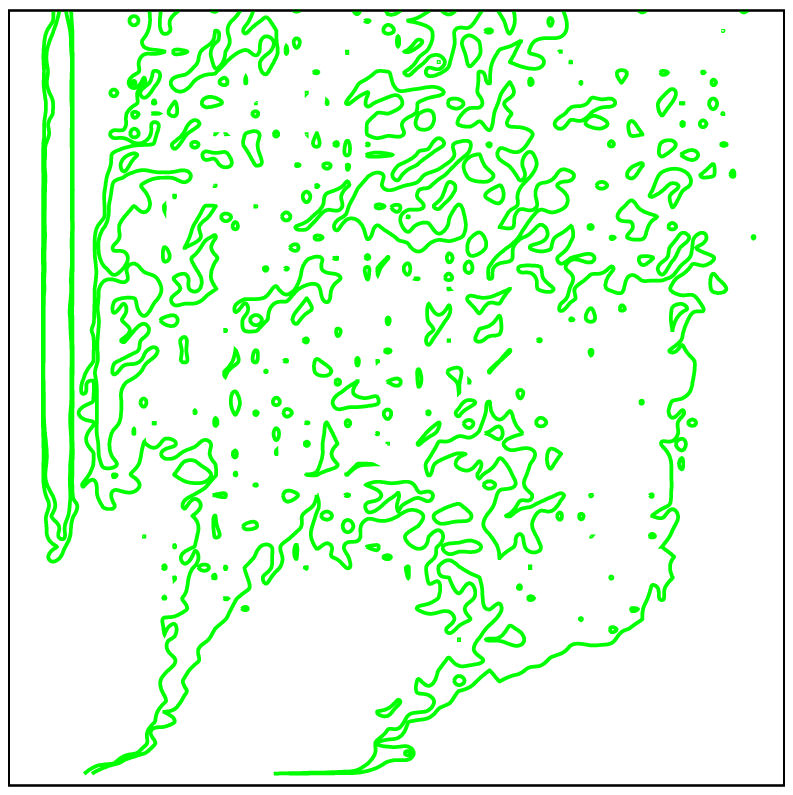}
\caption{Posterior probability distributions in the $m_0-m_{1/2}$ plane, taking into account the empirical inputs described in the text and using a natural prior (left) and a flat $\tan \beta$ prior (right), as described in the text. If naturalness considerations are taken into account, small $m_{1/2}$ and large $m_{0}$ are favored. In each frame, contours enclosing the 68\% and 95\% confidence regions are shown.
\label{2D}}}

In Fig.~\ref{3D}, we show the posterior probability
distribution marginalized to three dimensions, $m_0$, $m_{1/2}$
and $\tan \beta$, resulting from the fit. The darker inner surface contains
68$\%$ of the probability distribution and the outer lighter surface contains
95$\%$. In Fig.~\ref{2D}, we show the same distribution,
marginalized to two dimensions ($m_0$ and $m_{1/2}$).  

The shape of the posterior is dominated by the relic density
constraint: the CMSSM tends to give much too high values for $\Omega_{\chi} h^2$
in generic parts of parameter space unless there exists a specific mechanism
through which efficient annihilation can occur. 
On the left-hand side of Figs.~\ref{3D} and~\ref{2D} (natural priors), we see
that this results in  
a large posterior for the focus point region of parameter space, where 
there is a significant higgsino fraction in the composition of the lightest
neutralino, causing it to efficiently annihilate into fermion and/or gauge
boson pairs. There also exists a favored region in which $2m_{\chi_1^0}
\approx m_{h^0}$ at the lowest values of $m_{1/2}$, disconnected from the
other region. In this case, annihilation occurs through the lightest
CP-even Higgs resonance into $b$ and $\tau$ pairs. 

On the right-hand side of Figs.~\ref{3D} and~\ref{2D}, we show for comparison
the results found using the flat $\tan \beta$ prior. 
For low values of $\tan \beta$, we have a vertical 
funnel on the right hand side of Fig.~\ref{3D}, corresponding to the stau
co-annihilation region, where staus 
efficiently annihilate with the neutralino lightest SUSY particle (LSP)
because of quasi mass degeneracy. 
At high $\tan \beta$, but moderate values of $m_0$ and $m_{1/2}$,
$2m_{\chi_1^0} \sim m_{A^0}$, leading to efficient dark matter
annihilation through $s-$channel pseudo-scalar Higgs boson exchange, into $b$ and $\tau$ pairs. At low $m_{1/2}$, we again have the $h^0-$pole annihilation
region but, for flat $\tan \beta$ priors,
small values of $\tan \beta$ are disfavored as they lead to values
of $m_{h^0}$ which are below the LEP2 limit. The LEP2 Higgs mass constraint
also means that the $h^0$ region is outside the 68$\%$ contour. 
At high $m_0$, the focus point is also in evidence for flat $\tan \beta$
priors.  

\FIGURE{\includegraphics[width=2.7in,angle=0]{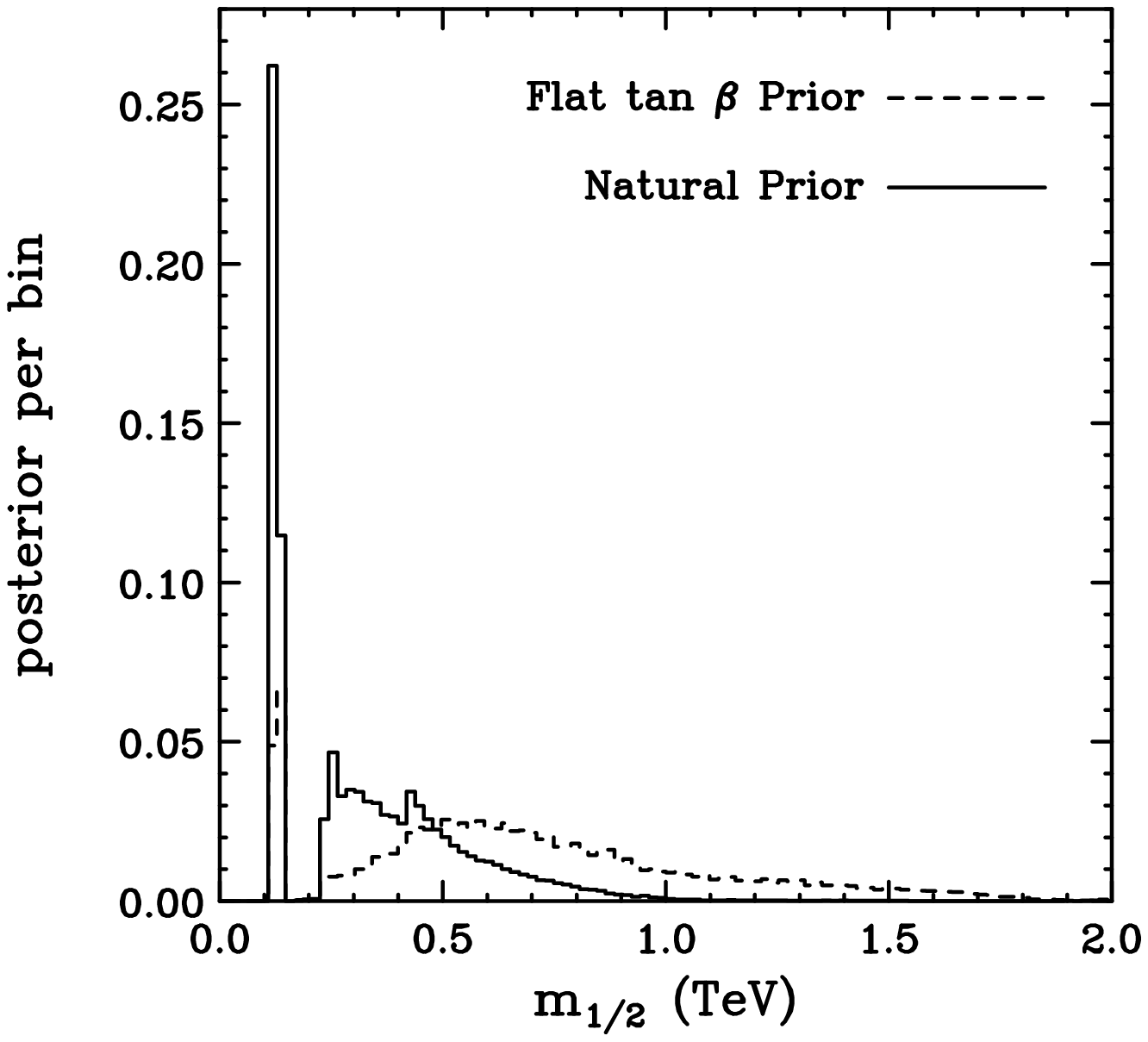}
\includegraphics[width=2.7in,angle=0]{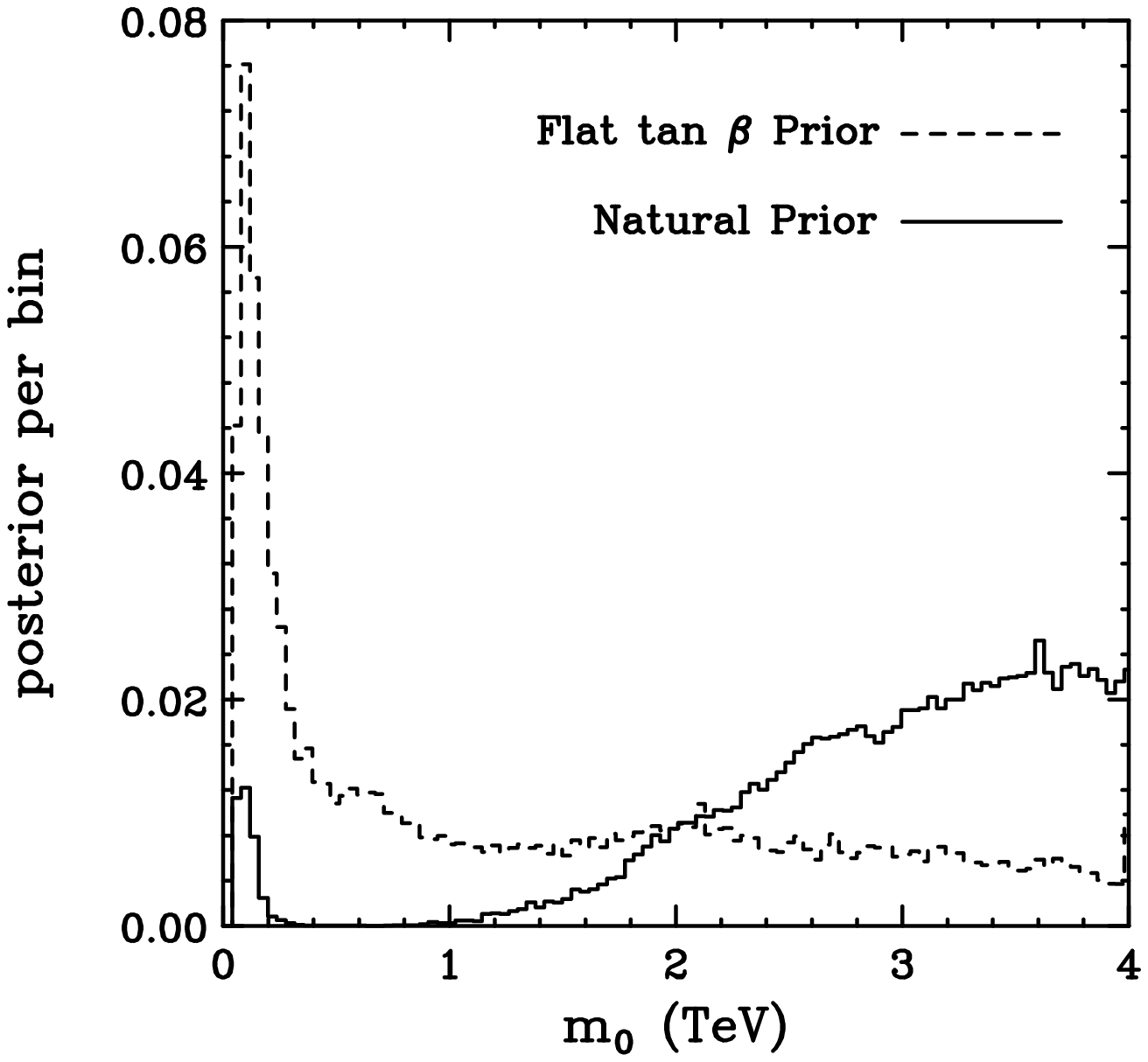}\\
\includegraphics[width=2.7in,angle=0]{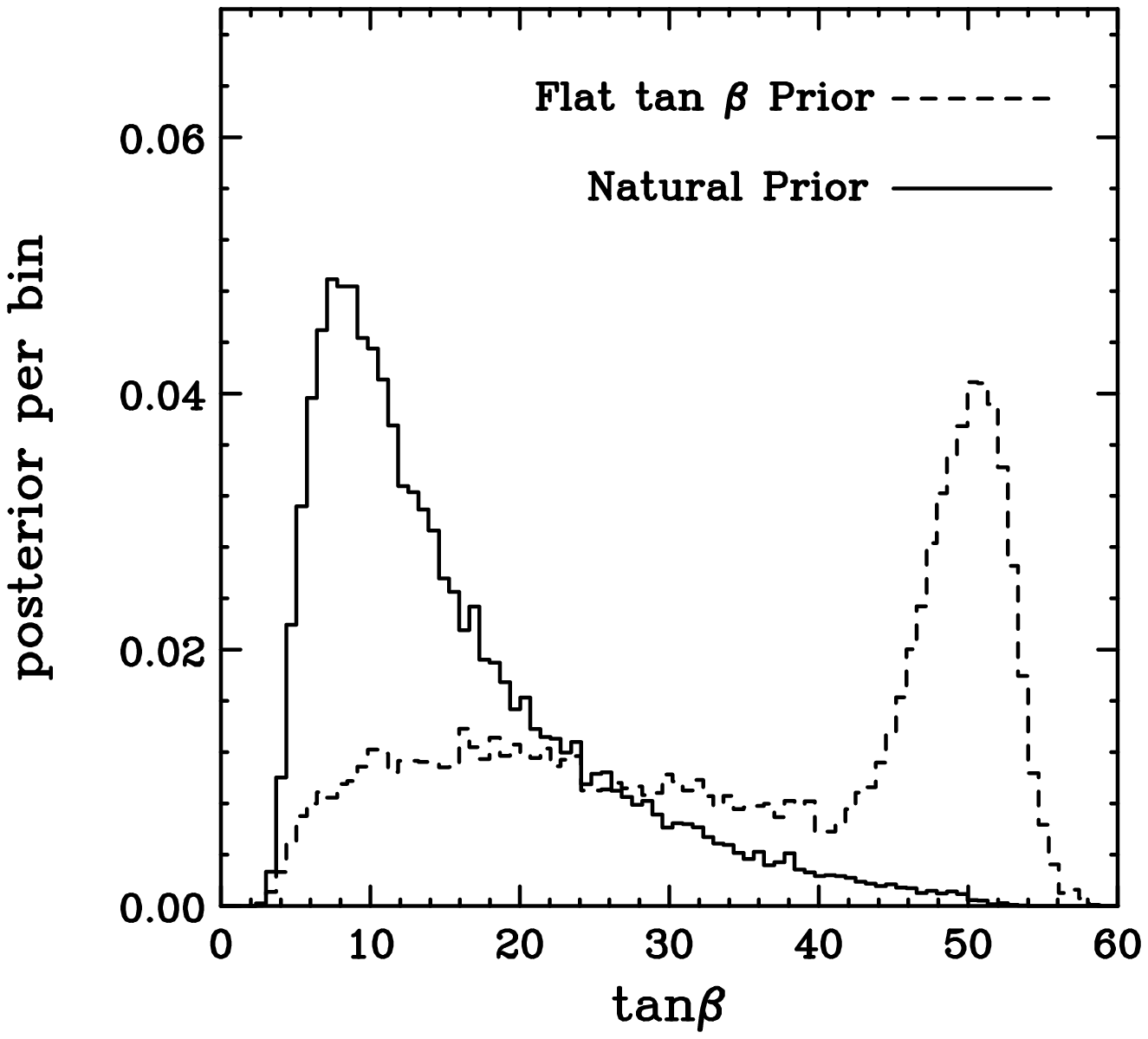}
\includegraphics[width=2.7in,angle=0]{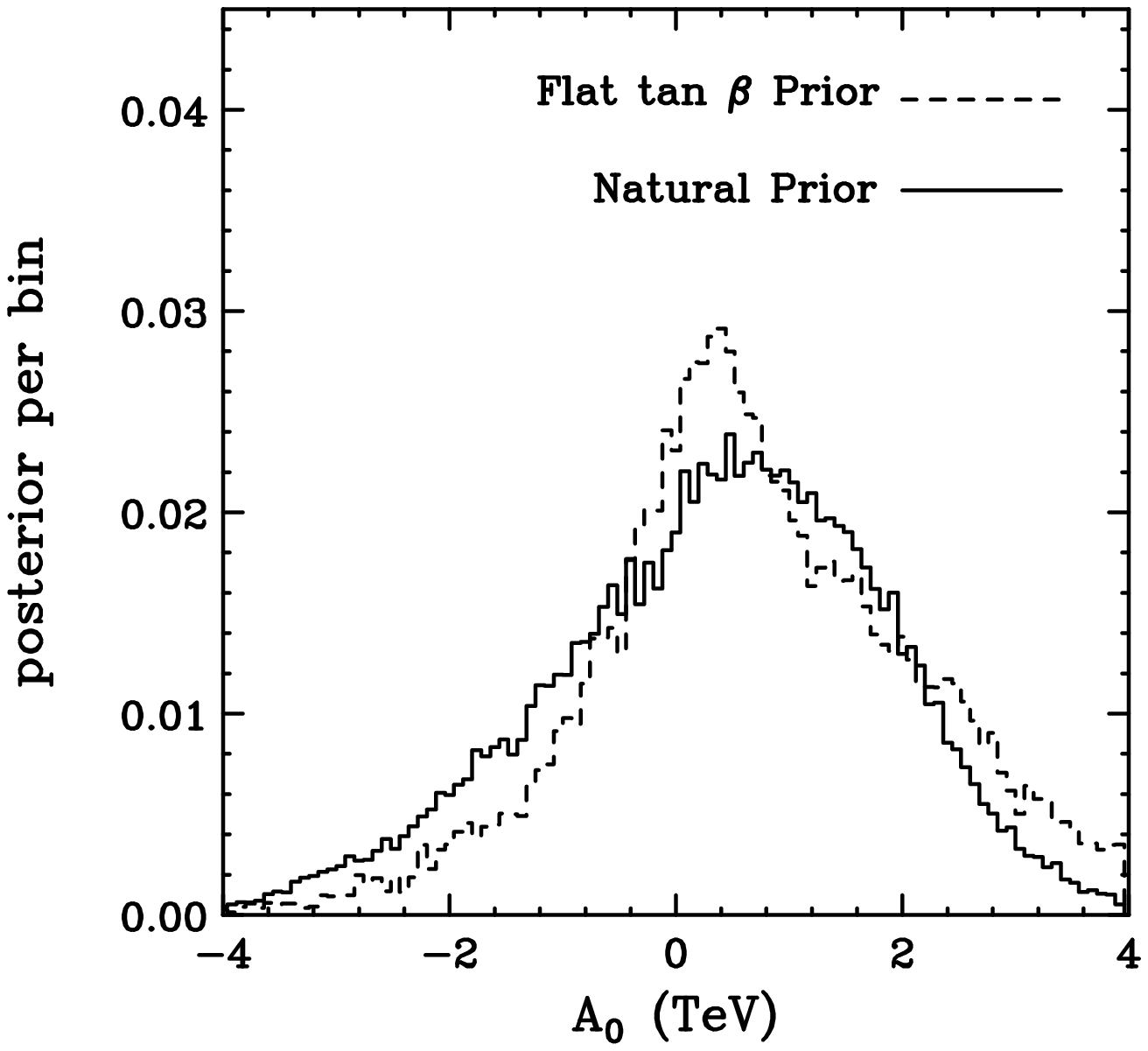}
\caption{Posterior probability distributions of the CMSSM parameters,
  marginalized over the unseen parameters, taking into account the empirical
  inputs described in the text and using a flat $\tan \beta$ prior 
  (dashed) or a natural prior
  (solid) as described in the text. For the natural measure, small to
  moderate values of $m_{1/2}$ and $\tan \beta$ 
  are preferred, while $m_{0}$ is strongly favored to be large. In each frame,
  each distribution is plotted in 100 bins of equal width.} 
\label{CMSSMvalues}
}
Some features of the posterior distribution become much clearer when
marginalized to one CMSSM parameter dimension.
Such marginalizations are shown in 
Fig.~\ref{CMSSMvalues}, one for each of the four continuous CMSSM
parameters. 
Three key differences are immediately noticed when comparing the results found
using the natural and flat priors. 
Firstly, in considerable contrast to the
flat $\tan \beta$ case, the natural prior strongly favors heavy sfermion masses (large
$m_0$). Secondly, the natural prior prefers low to moderate values of $\tan
\beta$, again in contrast to the flat $\tan \beta$ case. The posterior pdfs of
$m_0$ and $\tan \beta$ therefore display themselves to be strongly prior
dependent, whereas $m_{1/2}$ shows a smaller difference between the fits using
the two priors, and $A_0$ shows very little dependence. Thus, the choice of theoretical prejudice alters the results of the fit for
$\tan \beta$ and $m_0$.  
The natural
prior prefers somewhat
smaller values of $m_{1/2}$ relative to those found using the flat prior. 
From the $m_{1/2}$ figure, we see the strong bi-modality of the 
posterior, where the spike at low values of $m_{1/2}$ corresponds to the
$h^0-$pole region. 

Although one might expect values of $m_0$ much larger than $M_Z^{emp}$ to
require an unacceptable degree of fine tuning, this does not have to be the case. In particular, although large values of $m_0$ lead to large values of ${\bar
  m}_{H_{1,2}}^2$ unless counter-balanced by
an almost equally large value of $\mu^2$ in order to obtain the empirical value
of ${M_Z^2}^{emp} = (91$ GeV$)^2$, this fine tuning can be avoided in portions
of supersymmetric parameter space known as 
the focus point region (also known as the `hyperbolic branch'). In this region,
the RG trajectories of the 
Higgs mass parameters meet at a point near the weak scale, at which their
(small) values are independent of their input values at the UV boundary. This
leads to a 
Higgs potential which is largely insensitive to the scalar masses. As a
result, models with multi-TeV squarks, sleptons and heavy Higgs scalars can
exist with only a modest degree of fine tuning~\cite{feng}. 

\FIGURE{
\includegraphics[width=2.3in,angle=270]{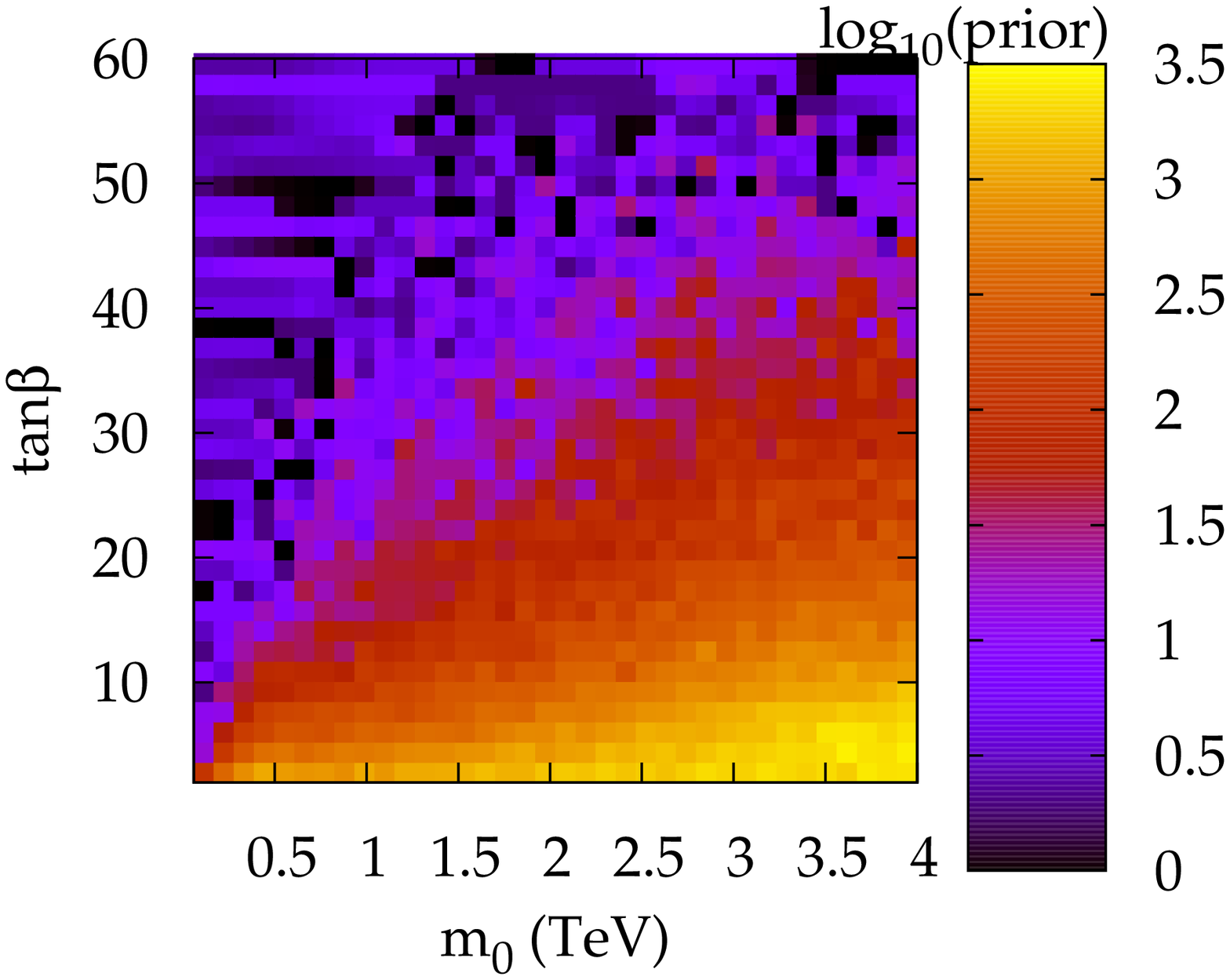}
\caption{Marginalised natural prior as a function of $m_0$ and $\tan \beta$.
 \label{fig:prior}}} 
Eq.~\ref{initial2} indicates that the natural 
prior favors lower $\tan \beta$ (thus suppressing the $A^0$ pole region) and
low values of $\mu$, which occur in the focus point region.  
We can see evidence
of the latter by examining Fig.~\ref{fig:prior}, where the logarithm of the
marginal prior probability density is plotted as a function of $m_0$ and $\tan
\beta$ for a scan where all data was ignored.
$\mu$ is
particularly low in the focus point region, and so the prior factor $1/\mu$ 
is the dominant factor in enhancing the focus point at high $m_0$. 
We also see the preference for lower values of $\tan \beta$ in the figure
evident in Eq.~\ref{initial2}. 
Ref.~\cite{Roszkowski:2007fd} assumed significantly smaller theoretical errors
on the prediction for the branching ratio of $b \rightarrow s \gamma$ than
ours. There, it was
found that the current best-predicted value of the branching ratio from the
Standard Model (found by including some higher order contributions in the
calculation), additional  
preference for the focus point region was found compared to the case where the
higher order contributions were neglected. If we were to reduce our assumed
theoretical errors on the prediction of this quantity, we would obtain a
similar further enhancement of the focus point region. 

A very distinctive dark matter phenomenology emerges in the majority of the
CMSSM parameter space favored by the natural prior. In particular,
most of this space contains a rather light neutralino, with a mixed
gaugino-higgsino composition as predicted by comparably low values of $\mu$
and $M_{1/2}$. Such mixed neutralinos, which appear in the
focus point region, have sizable couplings to SM gauge bosons and fermions
which enable them to annihilate efficiently and avoid being overproduced in
the early 
universe. There also exists a sizable probability (approximately \hpolePercentage\%),
however, for the parameter space in which the lightest neutralino falls in the
$h^0$-pole region, without a large degree of higgsino composition.
This region can be seen in the figures, and appears
at $m_{1/2}\sim 100$ GeV or $m_{\chi^0}\approx 60$ GeV.

Writing the lightest neutralino as a mixture of gauginos (bino and wino) and
higgsinos: 
\begin{equation}
\chi^0 = N_{11}\tilde{B}     +N_{12} \tilde{W}^3
          +N_{13}\tilde{H}_1 +N_{14} \tilde{H}_2,
\end{equation}
we define the gaugino and higgsino fractions as $|N_{11}|^2+|N_{12}|^2$ and
$|N_{13}|^2+|N_{14}|^2$, respectively. Within the CMSSM, the assumption that
the gaugino masses unify at a common scale ensures that $|N_{12}|^2$ is never
much larger than a few percent.  The relative bino and higgsino fractions of
the lightest neutralino are, therefore, largely dictated by the ratio of $M_1$
(determined by $m_{1/2})$ and $\mu$. 
\FIGURE{\includegraphics[width=2.7in,angle=0]{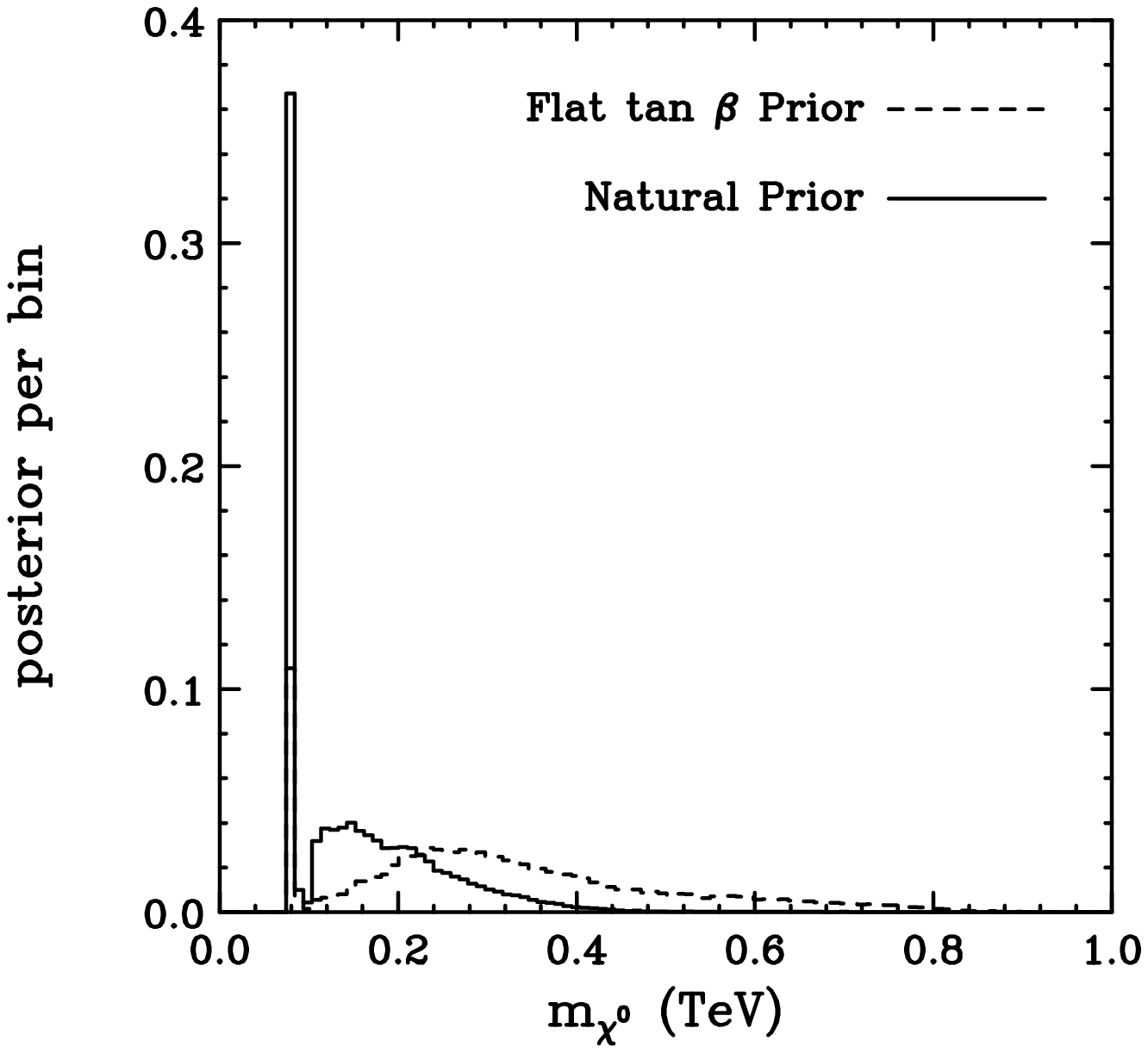}
\hfill
\includegraphics[width=2.7in,angle=0]{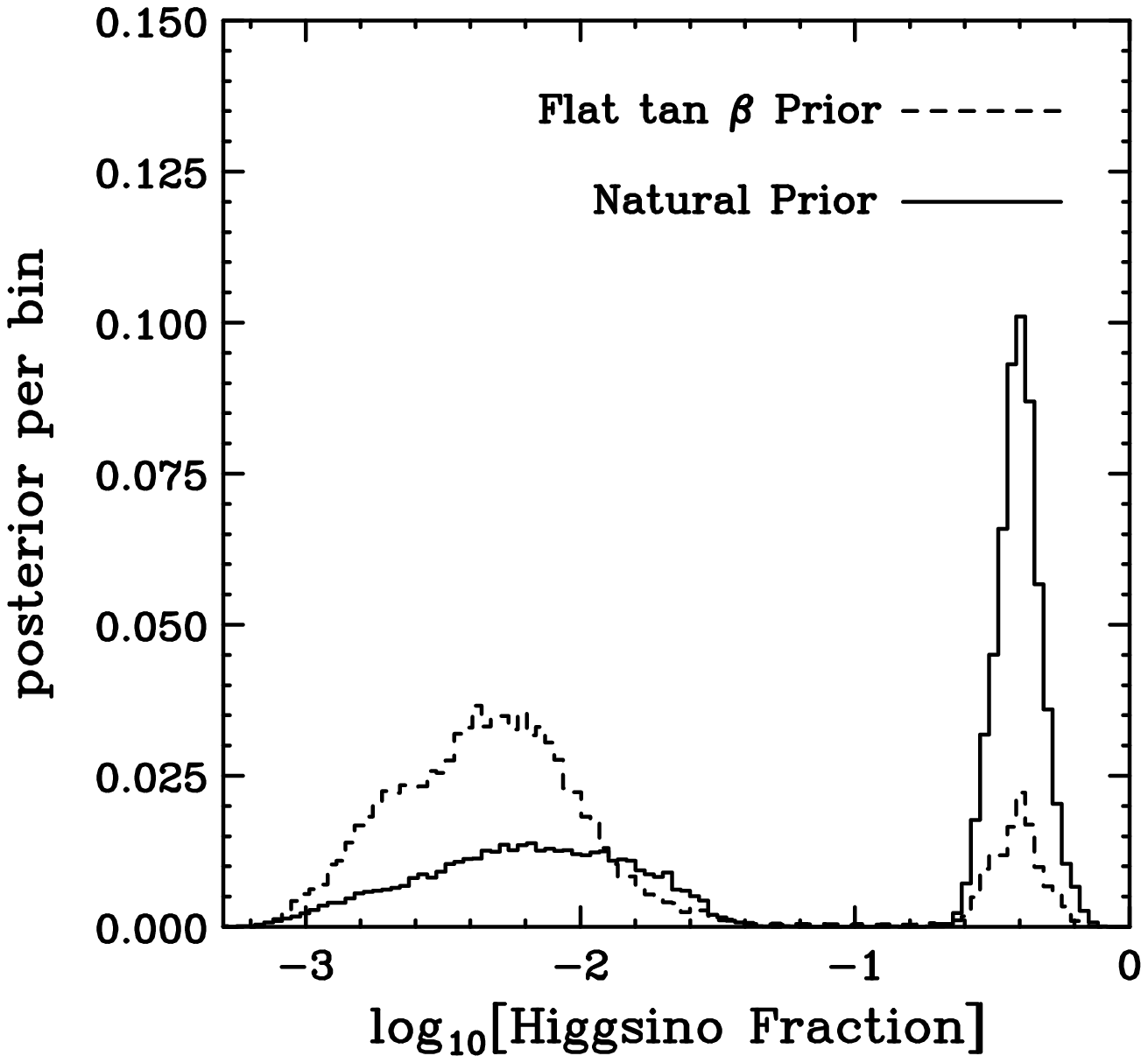}
\caption{Posterior probability distributions for the mass of the lightest
  neutralino and its higgsino fraction, using a   natural prior (solid) or a 
  flat $\tan \beta$ prior
  (dashed), as described in the text. If naturalness
  considerations are taken into account, a light neutralino with a mixed
  higgsino-gaugino composition is favored. In each frame, each distribution is
  plotted in 100 bins of equal width.} 
\label{lsp}}
In Fig.~\ref{lsp}, we show the posterior probability distributions for the
mass and higgsino fraction of the lightest neutralino. Interestingly, the
natural priors lead to a strong preference for a highly mixed
higgsino-bino composition for the lightest neutralino ($|N_{13}|^2+|N_{14}|^2
\sim 0.1$). This is a direct consequence of being in the focus point region of
supersymmetric parameter space. In particular, as $m_0$ is increased, the
value of $|\mu|$, as determined by the electroweak symmetry breaking
conditions, is driven to smaller values, thus increasing the higgsino content
of the lightest neutralino.  
In the following sections we will explore the phenomenology and detection
prospects for neutralino dark matter with these properties.  

\section{Direct Detection}

Searches for dark matter which attempt to detect such particles through their elastic scattering with nuclei are known as direct detection. Experiments currently involved in this effort include CDMS~\cite{cdms}, XENON~\cite{xenon},
ZEPLIN~\cite{zeplin}, Edelweiss~\cite{edelweiss}, CRESST~\cite{cresst},
WARP~\cite{warp}, KIMS~\cite{kims}, and COUPP~\cite{coupp}.

The ability of experiments such as these to detect a weakly interacting
massive particle (WIMP) depend on its mass
and on its elastic scattering cross section with the nuclei making up the
detector. The elastic scattering cross section of a neutralino or other WIMP
can be broken into spin-independent and spin-dependent
contributions. Spin-independent interactions represent coherent scattering
with the entire nucleus, and lead to a cross section proportional to the square of the target nucleus' mass. Spin dependent interactions, in contrast, lead to a cross section that scales with $J(J+1)$, where $J$ is the total spin of the target nucleus. Currently, direct constraints on spin-independent scattering are far more stringent than those for spin-dependent scattering. For this reason, we focus on spin-independent scattering in this section.

The spin-independent neutralino-nucleus elastic scattering cross section is given by:
\begin{equation}
\label{sig}
\sigma \approx \frac{4 m^2_{\chi^0} m^2_{T}}{\pi (m_{\chi^0}+m_T)^2} [Z f_p + (A-Z) f_n]^2,
\end{equation}
where $m_T$ is the mass of the target nucleus, and $Z$, $A$ are the atomic number and atomic mass of the nucleus, respectively.  $f_p$ and $f_n$
are the neutralino's couplings to protons and neutrons, given by~\cite{jungmanreview}:
\begin{equation}
f_{p,n}=\sum_{q=u,d,s} f^{(p,n)}_{T_q} a_q \frac{m_{p,n}}{m_q} + \frac{2}{27} f^{(p,n)}_{TG} \sum_{q=c,b,t} a_q  \frac{m_{p,n}}{m_q},
\label{feqn}
\end{equation}
where $a_q$ are the neutralino-quark couplings~\cite{jungmanreview,ascatter} and $f^{(p,n)}_{T_q}$ denote the quark content of the nucleon and have been measured to be:  $f^{(p)}_{T_u} \approx 0.020\pm0.004$,  
$f^{(p)}_{T_d} \approx 0.026\pm0.005$,  $f^{(p)}_{T_s} \approx 0.118\pm0.062$,  
$f^{(n)}_{T_u} \approx 0.014\pm0.003$,  $f^{(n)}_{T_d} \approx 0.036\pm0.008$ and 
$f^{(n)}_{T_s} \approx 0.118\pm0.062$~\cite{nuc}.  The first term in this equation corresponds to interactions with the quarks in the target, which can occur through either $t$-channel CP-even Higgs exchange, or $s$-channel squark exchange. The second term corresponds to interactions with gluons in the target through a quark/squark loop. $f^{(p)}_{TG}$ is given by $1 -f^{(p)}_{T_u}-f^{(p)}_{T_d}-f^{(p)}_{T_s} \approx 0.84$, and analogously, $f^{(n)}_{TG} \approx 0.83$. 

\FIGURE{\twographsCDMS{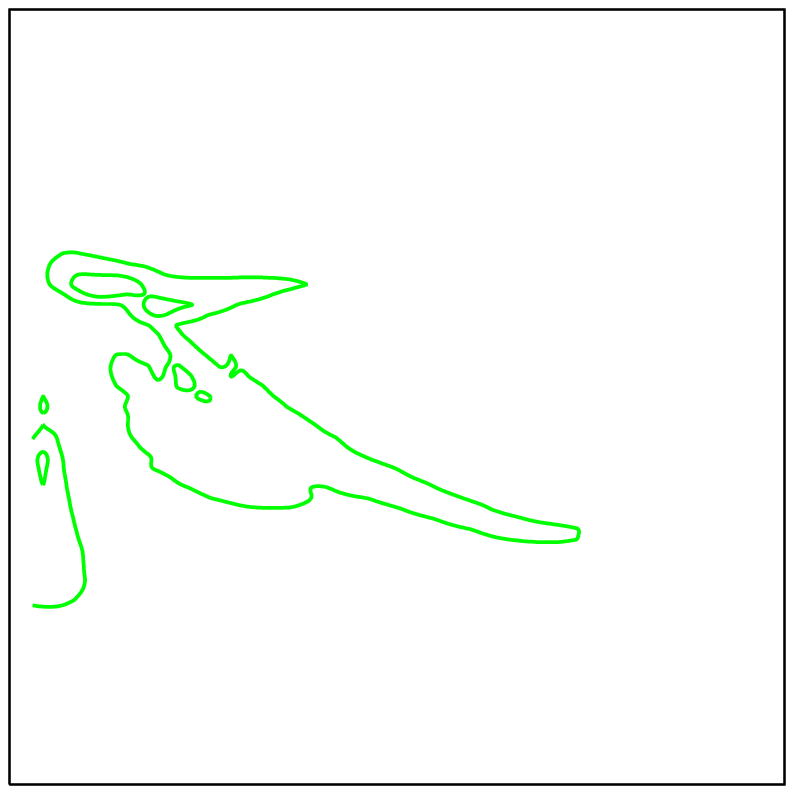}{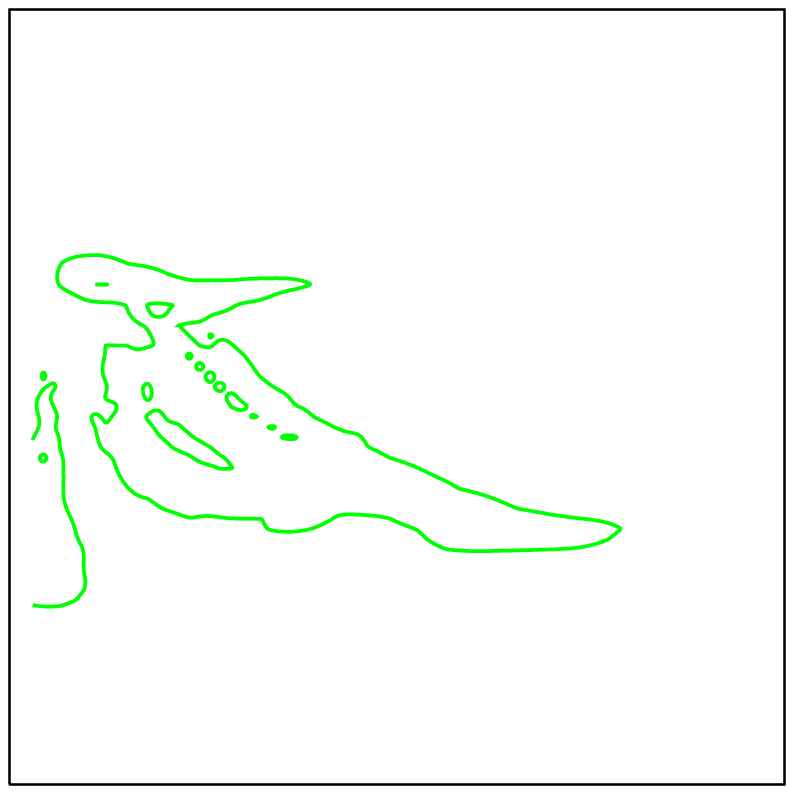}
\caption{Posterior probability distributions in the neutralino-nucleon,
  spin-independent elastic scattering cross section vs neutralino mass plane,
  taking into account the empirical inputs and using the 
  natural (left) and flat $\tan \beta$ (right) priors described
  in the text. If natural priors are used, the focus point region is preferred,
  leading to $\sigma_{\chi N, {\rm SI}} \sim 3 \times 10^{-8}$ pb. The light
  Higgs pole region is also seen in the left frame with $m_{\chi^0} \sim 60$
  GeV and a smaller cross section. In each frame, contours enclosing the 68\%
  and 95\% confidence regions are shown. Also shown is the 90$\%$ confidence
  level current upper bound
  placed by the CDMS collaboration~\cite{cdms} assuming a local dark matter
  density of $\rho_{\chi^0} = 0.3$ GeV/cm$^3$ and a 
characteristic velocity of $v_0 =230$ km/s.  
\label{SI}}}
In Fig.~\ref{SI}, we show the posterior probability distributions for the
neutralino's spin-independent elastic scattering cross section (per nucleon),
for the case of a natural prior (left) and a flat $\tan \beta$ prior
(right). From this 
figure, it is clear that the most probable parameter regions, corresponding to
a highly mixed neutralino in the focus point, are concentrated around
$\sigma_{\chi N, {\rm SI}} \sim 3 \times 10^{-8}$ pb, which is just beyond the
current reach of direct detection experiments such as CDMS~\cite{cdms}. It is
straightforward to see why is the case. In the focus point region of the
CMSSM, the squarks and heavy Higgs boson masses are large enough to not
contribute significantly to the process of neutralino elastic scattering. In
this case, and for a neutralino with a negligible wino content, the coupling
$a_q$ is proportional to $N_{11} N_{14}/m^2_h$. With the higgsino fraction predicted to be $\sim 0.1$ (as seen in the left frame of Fig.~\ref{lsp}), the resulting
elastic scattering cross section is expected to be quite large, leading to the results
found in the left frame of Fig.~\ref{SI}. 
The lower left portion of the left hand frame of Fig.~\ref{SI} corresponds
to the 
light Higgs pole region, in which the lightest neutralino is largely
bino-like. In this region, the neutralino-quark couplings and corresponding
cross sections with nuclei are smaller compared to the mixed bino-higgsino
region. In the right hand frame, we show the flat $\tan \beta$ prior direct
detection cross section  
posterior probability for comparison. Despite our updated constraints and
additional observables, the flat $\tan \beta$ posterior looks
indistinguishable to the eye to previous 
determinations~\cite{deAustri:2006pe,Roszkowski:2007fd}, where the connection
between a preference for the focus point and good direct detection prospects
were pointed out.

Before moving on to the prospects for indirect detection, a few comments are in order. Firstly, direct detection experiments do not simply measure the
WIMP's interaction cross section, but instead measure the cross section
multiplied by the 
flux of WIMPs passing through the detector. The observed rate, therefore, depends on the local density of dark matter and, to a lesser degree, on its velocity
distribution. Constraints such as those from CDMS shown in Fig.~\ref{SI} are
made under reasonable assumptions about the local dark matter density and
velocity distribution. Measurements of the Milky Way's rotation curves can be used to estimate a local dark matter density in the range of 0.22 to 0.73 GeV/cm$^3$~\cite{local}. As
long as the fine-grained structure of the dark matter distribution is not
highly clumpy, this range should be appropriate for the purposes of direct
detection (for discussions, see Ref.~\cite{fine}.) The nuclear
physics involved in neutralino-nuclei scattering also introduces a degree of
uncertainty into the constraints placed by direct detection experiments (for
more details, see Ref.~\cite{nuclear}).  

\section{Indirect Detection}

\subsection{Neutrino Telescopes}

Through elastic scattering with nuclei in the Sun, neutralinos can become
gravitationally bound, leading them to accumulate and annihilate in the Sun's
core. Such annihilations can potentially produce a flux of high energy
neutrinos observable to next generation neutrino
telescopes~\cite{neutrinos}.

\FIGURE{\twographs{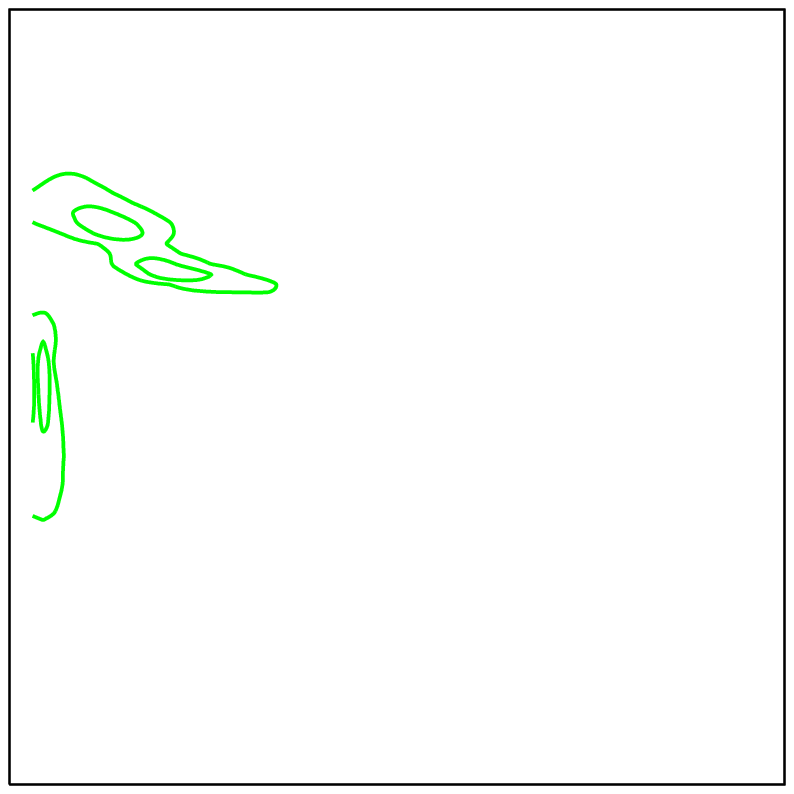}{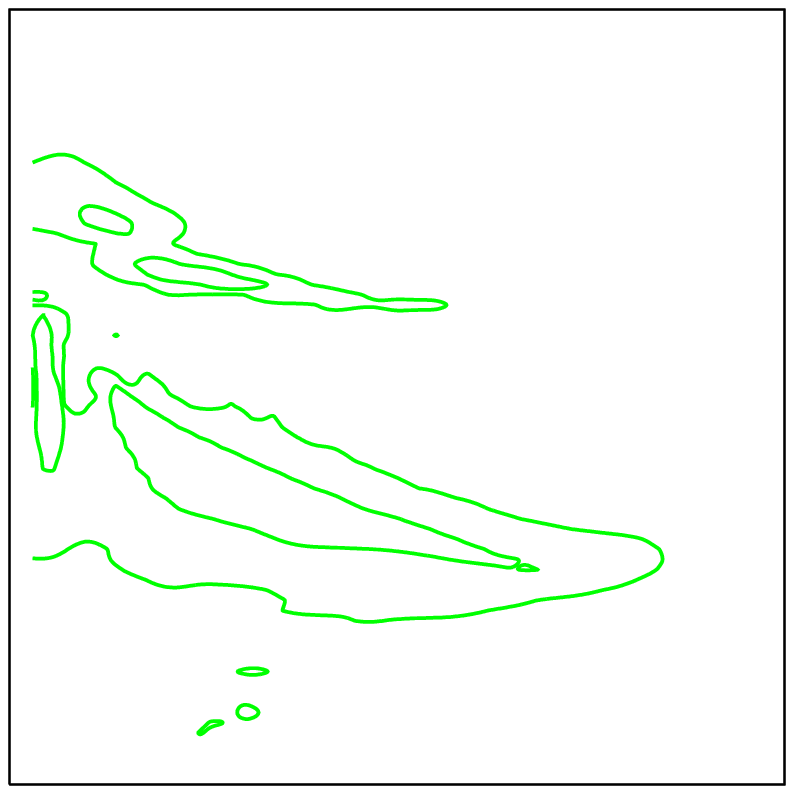}
\caption{Posterior probability distributions in the neutralino-proton,
  spin-dependent elastic scattering cross section vs neutralino mass plane,
  taking into account the empirical inputs and using the natural (left) and
  flat $\tan \beta$
  (right)  priors described in the text. If naturalness
  considerations are taken into account, the focus point region is preferred,
  leading to $\sigma_{\chi\mathrm{p, SD}} \sim 10^{-4}$ pb. The light Higgs
  pole region is also seen in the left frame with $m_{\chi^0} \sim 60$ GeV
  and a smaller cross section. In each frame, contours enclosing the 68\% and
  95\% confidence regions are shown. 
\label{SD}}}

Assuming a standard local density and velocity distribution, neutralinos become captured by the Sun at a rate given by~\cite{capture}: 
\begin{equation}
C^{\odot} \approx 3.35 \times 10^{19} \, \mathrm{s}^{-1} 
\left( \frac{\sigma_{\chi \mathrm{p, SD}} +\, \sigma_{\chi \mathrm{p, SI}}
+ 0.07 \, \sigma_{\chi \mathrm{He, SI}}     } {10^{-7}\, \mathrm{pb}} \right)
\left( \frac{100 \, \mathrm{GeV}}{m_{\chi^0}} \right)^2 ,
\label{capture}
\end{equation}
where $\sigma_{\chi \mathrm{p, SD}}$, $\sigma_{\chi \mathrm{p, SI}}$ and
$\sigma_{\chi \mathrm{He, SI}}$ are the spin dependent (SD) and spin
independent (SI) elastic scattering cross sections of neutralinos with
hydrogen (protons) and helium nuclei, respectively. The factor of $0.07$
reflects the solar abundance of helium relative to hydrogen and well as
dynamical factors and form factor suppression.

In the previous section, we calculated the posterior probability for the
neutralino's spin-independent elastic scattering cross section. In
Fig.~\ref{SD}, we show the analogous result for the spin-dependent,
neutralino-proton elastic scattering cross section. Again, we find that the
natural priors lead to a strong preference for large elastic
scattering cross sections. In this case, this results from the sizable
higgsino couplings to the $Z$-boson, which leads to a cross section which
scales as: $\sigma_{\chi^0 \mathrm{p, SD}} \propto
[|N_{13}|^2-|N_{14}|^2]^2$. By comparing Figs.~\ref{SI} and~\ref{SD}, we
clearly see that the spin-dependent cross section will dominate the overall
capture rate of neutralinos in the Sun. The flat $\tan \beta$ prior frame on
the right hand side looks rather similar to the posterior obtained recently in
the literature~\cite{Roszkowski:2007fd}, despite the fact that it has been
obtained with updated data and additional observables.

If the capture rate and annihilation cross section are sufficiently large,
equilibrium will be reached between these processes.   
For a number of neutralinos in the Sun, $N$, the rate of change of this
quantity is given by:
\begin{equation}
\dot{N} = C^{\odot} - A^{\odot} N^2, \nonumber
\end{equation}
where $C^{\odot}$ is the capture rate and $A^{\odot}$ is the 
annihilation cross section times the relative neutralino velocity per unit
volume. The present neutralino annihilation rate in the Sun is given by:
\begin{equation} 
\Gamma = \frac{1}{2} A^{\odot} N^2 = \frac{1}{2} \, C^{\odot} \, 
\tanh^2 \left( \sqrt{C^{\odot} A^{\odot}} \, t_{\odot} \right) \nonumber 
\end{equation}
where $t_{\odot} \approx 4.5$ billion years is the age of the solar system.
The annihilation rate is maximized when it reaches equilibrium with
the capture rate ({\it ie.}\/ when $\sqrt{C^{\odot} A^{\odot}} t_{\odot} \gg 1$). For the vast majority of the favored parameter space, we find that this condition is easily satisfied.

Neutralinos can generate neutrinos through a wide range of annihilation channels. Annihilations to heavy quarks, tau leptons, gauge bosons and Higgs bosons can each generate neutrinos in their subsequent fragmentation and decay. The muon neutrino spectrum at the Earth from neutralino annihilations in the Sun is given by:
\begin{equation}
\frac{dN_{\nu_{\mu}}}{dE_{\nu_{\mu}}} = \frac{C_{\odot} F_{\rm{Eq}}}{4 \pi
  D_{\rm{ES}}^2}   \bigg(\frac{dN_{\nu}}{dE_{\nu}}\bigg)^{\rm{Inj}}, 
\label{wimpflux}
\end{equation}
where $C_{\odot}$ is the capture rate of neutralinos in the Sun, $F_{\rm{Eq}}$ is the non-equilibrium suppression factor ($\approx 1$ for capture-annihilation equilibrium), $D_{\rm{ES}}$ is the Earth-Sun distance and $(\frac{dN_{\nu}}{dE_{\nu}})^{\rm{Inj}}$ is the neutrino spectrum from the Sun per neutralino annihilating. Due to $\nu_{\mu}-\nu_{\tau}$ vacuum oscillations, the muon neutrino flux observed at Earth is the average of the $\nu_{\mu}$ and $\nu_{\tau}$ components. 

Muon neutrinos produce muons in charged current interactions with nuclei in the material inside or near the detector volume of a high energy neutrino telescope. The rate of neutrino-induced muons observed in a high-energy neutrino telescope is given by: 
\begin{equation}
N_{\rm{events}} \approx \int \int \frac{dN_{\nu_{\mu}}}{dE_{\nu_{\mu}}}\, \frac{d\sigma_{\nu}}{dy}(E_{\nu_{\mu}},y) \,R_{\mu}((1-y)\,E_{\nu})\, A_{\rm{eff}} \, dE_{\nu_{\mu}} \, dy,
\end{equation}
where $\sigma_{\nu}(E_{\nu_{\mu}})$ is the neutrino-nucleon charged current
interaction cross section, $(1-y)$ is the fraction of neutrino energy which
goes into the muon and $A_{\rm{eff}}$ is the effective area of the
detector. $R_{\mu}$ is either the distance a muon of energy,
$E_{\mu}=(1-y)\,E_{\nu}$, travels before falling below the muon energy
threshold of the experiment, called the muon range, or the width of the
detector, whichever is larger.  The spectrum and flux of neutrinos generated
in neutralino annihilations is determined by its mass and dominant
annihilation modes.

\FIGURE{\includegraphics[width=2.7in,angle=0]{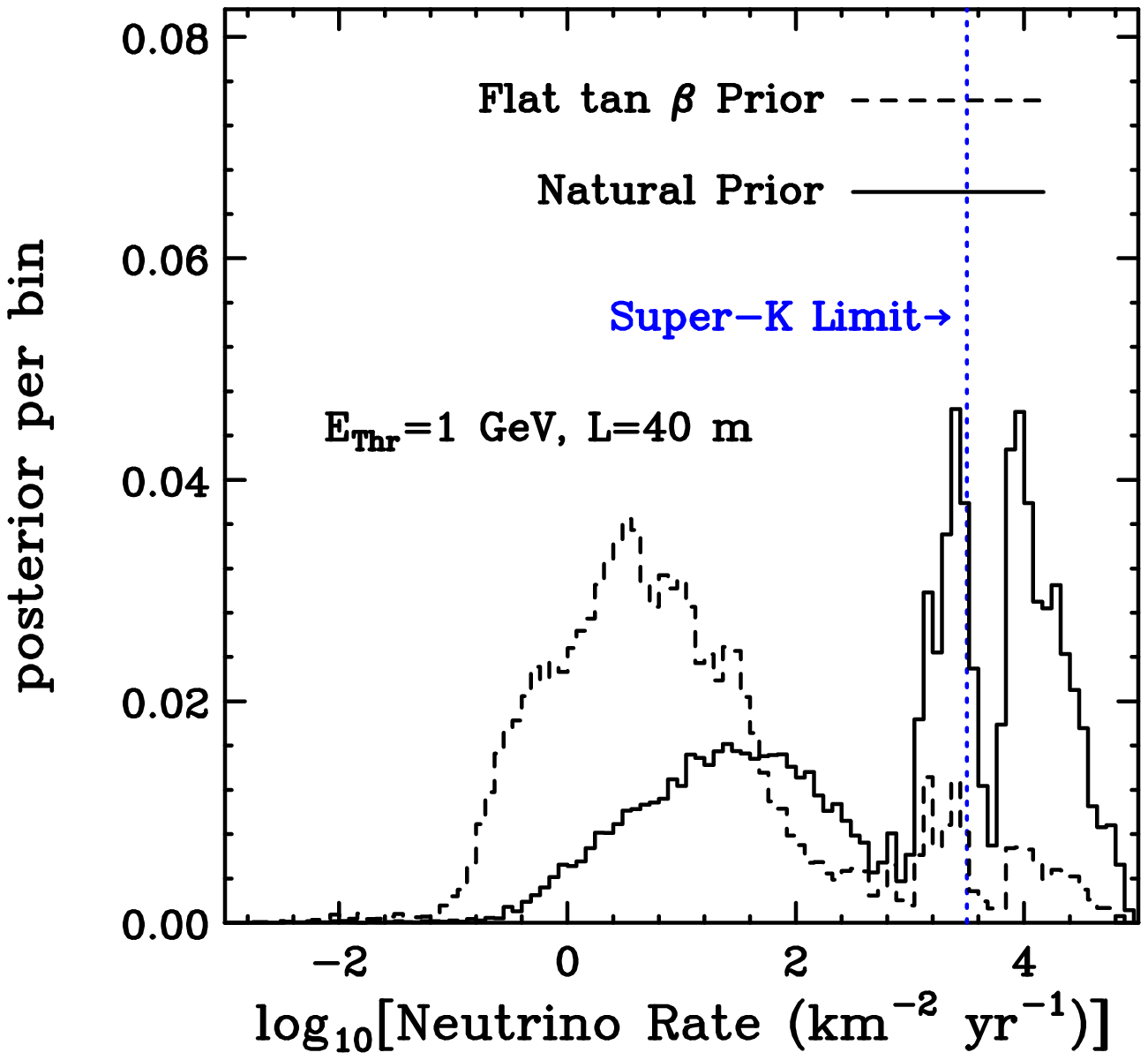}
\hfill
\includegraphics[width=2.7in,angle=0]{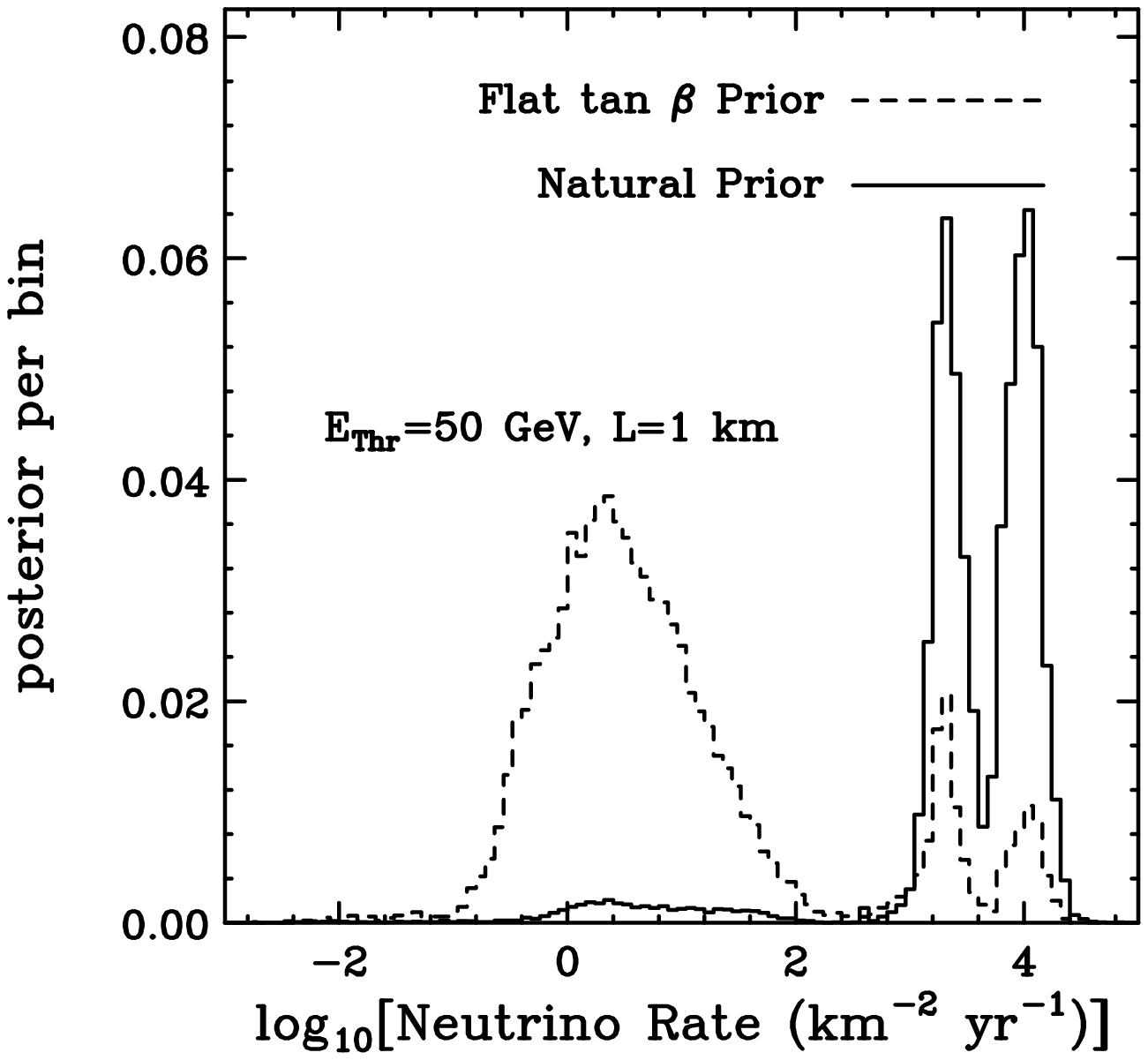}
\caption{Posterior probability distributions for the rate of neutrino events
  from neutralino annihilations in the Sun (per square kilometer, per year),
  using a flat $\tan \beta$ prior (dashed) or a natural prior (solid) as
  described in the text. The left (right) frame corresponds to the rate predicted at the Super-Kamiokande (IceCube) experiment. In
  each frame, each distribution is plotted in 100 bins of equal width. Note
  that in the right frame, \nonFPPercentage\% of the distribution 
  (the union of the light Higgs-pole
  region and the stau co-annihilation region) does not appear, as no events
  above the 50 GeV threshold are 
  generated.} 
\label{neutrino}
}

In Fig.~\ref{neutrino}, we show the posterior probability distribution for the
rate of neutrino induced muons from dark matter annihilations in the Sun in a
experiment such as Super-Kamiokande (left)~\cite{superk} and in a
kilometer-scale, high energy neutrino telescope such as IceCube
(right)~\cite{icecube}. For Super-Kamiokande, we plot the rate of muons with
an energy greater than 1 GeV, and use a detector width of 40 meters. For the
case of IceCube, we have used a 50 GeV muon energy threshold, and a kilometer
width.  

Currently, the strongest constraint on the neutrino flux from dark matter
annihilations in the Sun comes from Super-Kamiokande, which has placed an
upper limit on the rate of neutrino-induced muons from the Sun of
approximately $3 \times 10^3$ per square kilometer, per
year~\cite{superk}. Slightly weaker constraints have also been placed by
Amanda~\cite{amanda}, Baksan~\cite{baksan} and Macro~\cite{macro}. The
approximate Super-Kamiokande constraint is shown as a vertical dotted line in
the left frame of Fig.~\ref{neutrino}. Assuming an average local dark matter density of 0.3 GeV/cm$^3$, this bound excludes a sizable
fraction (38\%) of the probability distribution favored by our analysis. If a rather conservative value of 0.1 GeV/cm$^3$ were used instead, only 22\% of the of the probability distribution is excluded by the Super-Kamiokande limit.

The predicted rates in IceCube, as shown in the right frame of
Fig.~\ref{neutrino}, are extremely promising. About \fpPercentage\% of the
probability 
distribution corresponds to models which would produce thousands of muon induced
neutrino events per year from dark matter annihilations in the Sun. In
contrast, the rate of atmospheric neutrino induced muons in the same angular
window is only approximately 500 events per square kilometer per
year. Therefore, on the order of only $5\times \sqrt{500} \sim 100$ events per
square kilometer, per year would be required to produce a $5\sigma$
detection in IceCube. 

The \hpolePercentage\% of the probability distribution that falls in the light
Higgs-pole 
region cannot be easily observed by IceCube, however, as these models contain
neutralinos with $\approx 60$ GeV masses, well below the range required to
generate muons above IceCube's energy threshold.  


\subsection{Gamma-Rays and Charged Particles}

Dark matter annihilating throughout the Milky Way's halo can potentially lead
to observable fluxes of gamma-rays, electrons/positrons, antiprotons and/or
antideuterons. The strategies for detecting gamma-rays from dark matter
annihilation are quite different from those for charged particle searches, as
gamma-rays travel undeflected by magnetic fields, making the observation of
point-like or extended regions of high dark matter density possible. Some of
the most promising regions include the center of the Milky Way~\cite{gc} and
nearby dwarf satellite galaxies~\cite{dwarf}. Charged particles produced in
dark matter annihilations, in contrast, diffuse in the galactic magnetic field
erasing any directional information. Nonetheless, if the rate of dark matter
annihilation is large enough in the galactic halo, it may be possible to
identify its contribution in the antimatter component of the cosmic ray
spectrum. Additionally, electrons and positrons produced through dark matter
annihilations could potentially produce an observable flux of synchrotron
radiation as they travelling through the Galactic Magnetic Fields~\cite{syn}.

\FIGURE{\twographsGLAST{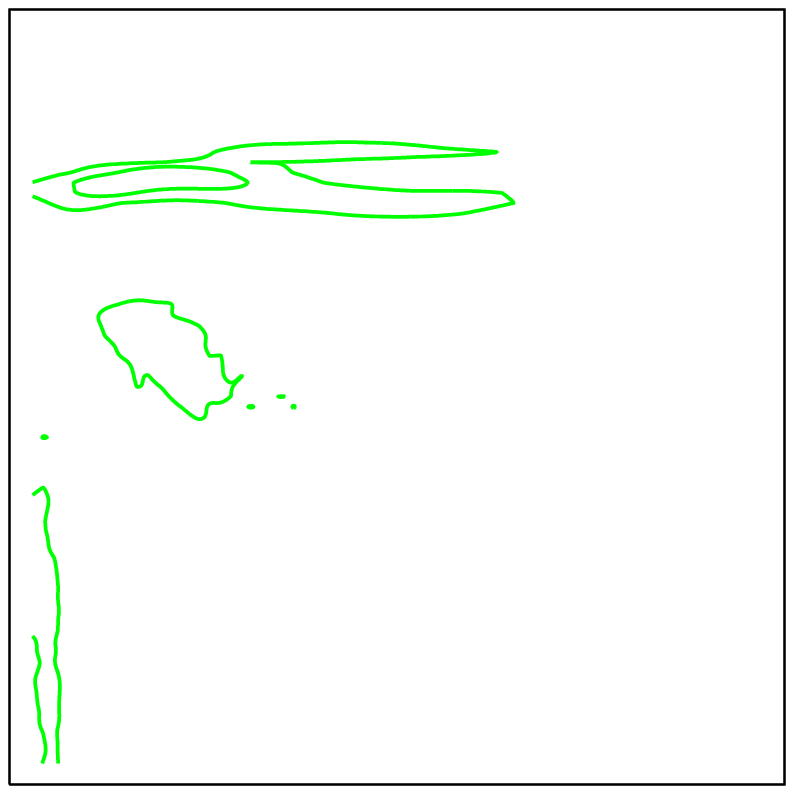}{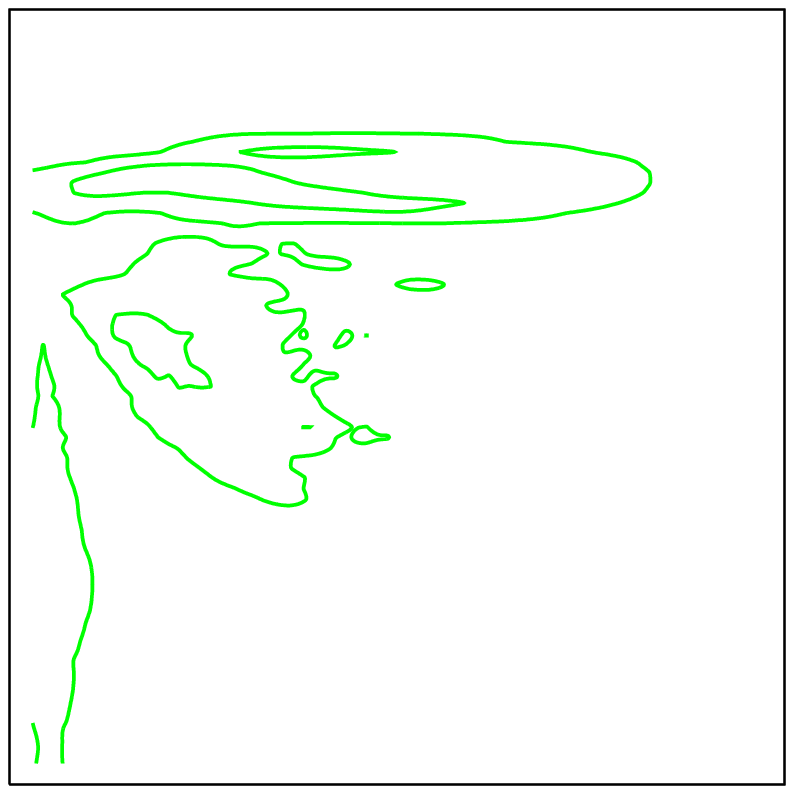}
\caption{Posterior probability distributions in the annihilation cross section
  vs lightest neutralino mass plane, taking into account the empirical inputs
  and using 
  the natural (left) and flat $\tan \beta$ (right) priors described in
  the text. If naturalness considerations are taken into account, the focus
  point region is preferred, leading to $\sigma_{\rm Ann} v \sim 3 \times
  10^{-26}$ cm$^3$/s. The light Higgs pole region is also seen in the left
  frame with $m_{\chi^0} \sim 60$ GeV and a smaller cross section times
  relative velocity. In each
  frame, contours enclosing the 68\% and 95\% confidence regions are
  shown. Also shown is the reach of the GLAST telescope for the case of a
  Navarro, Frenk and White (NFW), halo profile~\cite{dodelson}. 
\label{sigmav}}}

The prospects for detecting dark matter with gamma-rays and charged particles each
depend on both particle physics and astrophysical inputs. Regarding particle
physics, the neutralino's annihilation cross section and mass (and to a lesser
extent its dominant annihilation modes) each impact the reach of indirect
detection efforts. In Fig.~\ref{sigmav}, we show the posterior probability
distribution for the annihilation cross section and mass of the lightest
neutralino. Using natural priors, the favored focus point region leads to a cross section times 
relative velocity of $\sigma_{\rm Ann} v \approx 3 \times 10^{-26}$ cm$^3$/s,
which is 
approximately the maximal value possible for a thermal WIMP\@. The models in the
light Higgs-pole region have considerably smaller annihilation cross section,
making their indirect detection with gamma-rays or charged cosmic rays very
unlikely. If flat $\tan \beta$ priors are used, the neutralino's annihilation
cross section can be considerably smaller than the bulk of the favored natural
prior region.

A number of astrophysical inputs also impact the reach of gamma-ray,
cosmic ray and synchrotron searches for dark matter annihilation. In the case
of gamma-rays and synchrotron radiation, the 
annihilation rate in the inner galaxy or elsewhere depends on the integral
of the dark matter density squared, over the observed line-of-sight. This
leads to a strong dependence on the density of dark matter in the inner
parsecs of halos, well beyond the resolution of current N-body
simulations. Furthermore, the gravitational potential in the inner region of
the Milky Way is dominated by baryons rather than dark matter, whose effects
are not generally included in such simulations. Although the impact of baryonic matter
on the dark matter distribution is difficult to predict, an enhancement in the
dark matter density and corresponding annihilation rate is expected to result
from the process of adiabatic compression~\cite{ac}. The adiabatic accretion
of dark matter onto the central super-massive black hole might also lead to the formation of a density spike in the dark matter
distribution, leading to an enhanced dark matter annihilation
rate~\cite{spike}. Collectively, these astrophysical uncertainties lead to
several orders of magnitude of variation in predictions of the gamma-ray flux
from dark matter annihilations.  

In Fig.~\ref{sigmav}, we show the projected reach of the GLAST gamma-ray
telescope~\cite{glast} after ten years of observation, as calculated in Ref.~\cite{dodelson}, for the case of
a dark matter distribution following the Navarro, Frenk and White (NFW)
profile~\cite{nfw}, neglecting adiabatic compression and any other such effects. For this choice of the dark matter distribution, a non-negligible fraction of the posterior probability distribution favored by the natural priors are within GLAST's reach. One again, we remind the reader that variations from the NFW profile could modify this projection considerably. If the dark matter annihilation rate is even mildly enhanced from that predicted for a simple NFW profile, GLAST could potentially probe the entire range of CMSSM parameter space favored by the natural priors. This is in contrast to the results found using flat $\tan \beta$ priors, which allow for neutralinos with much smaller annihilation cross sections (see also Ref.~\cite{Roszkowski:2007va}). We also note that, if the lightest
neutralino is heavier than a few hundred GeV, ground-based atmospheric
Cerenkov telescopes could also be used to search for dark matter annihilations
in the inner Milky Way and elsewhere~\cite{act}. 

The prospects for the detection of cosmic ray electrons/positrons, antiprotons and
antideuterons are also subject to large degree of astrophysical
uncertainty. Namely, the diffusion parameters that describe the magnetic and
radiation fields of the Milky Way~\cite{diffusion}, as well as the dark matter
distribution, can significantly impact the reach of experiments such as
PAMELA~\cite{pamela} and AMS-02~\cite{ams02} to detect the presence of dark
matter annihilations. In particular, the ``boost factor'' that results from inhomogeneities in the dark matter distribution can
impact the prospects for such experiments considerably. For moderate choices
of the diffusion parameters and boost factor ($\sim$1-10), prospects for PAMELA
to detect positrons from dark matter annihilations over the background of
cosmic ray secondaries are similar to those for GLAST to detect
gamma-rays~\cite{silkpos}, as shown in Fig.~\ref{sigmav}.

\section{Summary and Conclusions \label{sec:conc}}

By considering measurements of quantities such as the anomalous magnetic
moment of the 
muon, the $b \rightarrow s \gamma$ branching fraction, the $B_s \rightarrow
\mu^+  \mu^-$ branching fraction, the mass of the $W$ boson, the effective
leptonic mixing angle, Higgs boson and sparticle search constraints, and the
cosmological dark matter abundance, it is possible to constrain the parameter
space of supersymmetry. The results of global fits to such indirect data
currently depend, however, on the choice of priors which are adopted. In most
of the 
global fits of supersymmetric parameter space which have been performed to
date, priors have been used which are flat in the derived quantity, $\tan
\beta$. A far more natural choice would be to use priors which are flat 
(or perhaps, logarithmic) in the
fundamental parameters $\mu$ and $B$. In this article, we have considered the
impact of adopting such natural priors upon fits of the CMSSM to indirect data,
focusing on the phenomenology of neutralino dark matter that is found in the
parameter space favored by such fits. 

Using natural priors and updated indirect data, we find a that two regions of
the CMSSM parameter space are strongly favored. Firstly, about
\fpPercentage\% of the 
posterior probability distribution corresponds to the focus point region. Of
the 
remainder, \hpolePercentage\% of the posterior probability distribution
corresponds to the 
light Higgs-pole region in which the lightest neutralino annihilates on
resonance with the light Higgs boson and \stauCoanPercentage\% corresponds to
the stau co-annihilation region, where staus and other sleptons efficiently
annihilate with the lightest neutralinos. In contrast to the results found
using 
priors flat in $\tan \beta$, we find that the stau-co-annihilation and
$A$-funnel regions of 
the CMSSM parameter space contribute negligibly to the posterior probability
distribution. 

In the favored focus point region, the lightest neutralino is a mixed
gaugino-higgsino ($\sim$10\% higgsino fraction) with a mass less than
approximately 300 GeV. Such a neutralino has a very distinctive dark matter
phenomenology and is nearly optimally suited for the purposes of direct and
indirect detection. In particular, a 
mixed gaugino-higgsino neutralino possesses
large couplings to Standard Model fermions, and thus has large elastic
scattering 
cross sections with nuclei. In the light Higgs-pole region, the lightest
neutralino can have considerably smaller couplings. 

We find that neutralinos in the favored focus point region have a
spin-independent elastic scattering cross section with nucleons of $\sim 3
\times 10^{-8}$ pb, which is within a factor of 2 (5) of the current limit
from CDMS for a 100 GeV (300 GeV) neutralino. We, therefore, expect direct
detection experiments to probe the majority of the posterior probability
distribution of the CMSSM parameter space in the very near future.  

The prospects for neutrino telescopes found using natural priors are also
very promising. In particular, most of the favored parameter space predicts
thousands of events to be observed per year in a kilometer-scale neutrino
telescope such as IceCube. Current constraints from Super-Kamiokande already
exclude 38\% of the posterior probability distribution, assuming a local dark
matter density of 0.3 GeV/cm$^3$.  

Although searches for dark matter using gamma-rays and charged particles
depend strongly on unknown astrophysical inputs, our analysis finds that the
majority of the favored parameter space predicts a neutralino annihilation
cross section near the maximum possible for a thermal relic ($\sim 3 \times
10^{-26}$ cm$^3$/s). This along with the relatively light mass range favored
for the lightest neutralino makes the prospects for GLAST and PAMELA to detect
neutralino dark matter near optimal.  

We believe that a prior that is flat in $\mu, B$ is a much more natural choice
than one flat 
in $\tan \beta$.If
the naturalness prior were complemented with an additional hyper-parameter prior
that enforces that all soft terms are ``of the same order''~\cite{weather}, 
the focus point is {\em dis}favoured. 
However one can consider
differences in the derived posterior probability distributions from the
different priors
as evidence that more data is needed to constrain the model. Thus, fit
predictions that are robust (i.e. approximately invariant) with respect to 
changes in assumed prior distributions are not attained for mSUGRA, since it
has many parameters and the data constraining it are rather indirect. 
While this is
undeniably true, it is still interesting to examine the effect of the more
natural prior as it gives us our ``best bet'' for quantities such as the 
dark matter-nucleon cross-sections relevant for direct detection, or
galactic annihilation cross sections relevant for indirect detection. 
Our neutralino-nucleon cross-sections coming from the fit for the 
flat $\tan \beta$ prior are similar to other previous determinations in the
literature, providing validation of our calculations. 
Our best guess for this quantity leads to a good chance that a further
increase of a factor of 10 in sensitivity by the experiments will lead to 
a direct detection discovery.
Had, instead of priors flat in $\mu, B$, we had chosen priors
that are flat in $\log \mu$ and $\log B$, we expect that our fits would
show even stronger preference for the focus point: 
an additional factor of $1/(\mu B)$ 
in the integrand of Eq.~\ref{initial2} would have lead 
to even more preference for the focus point region, with an associated extra
boost in detection cross-sections. 


Taken together, the results presented in this article are very encouraging for
the prospects for direct and indirect efforts to detect neutralino dark
matter. If natural choices are made in constructing priors, fits to the
currently available data predict that, within the context of the CMSSM, the
lightest neutralino is likely to have large elastic scattering and
annihilation cross sections. In particular, the majority of the posterior
probability distribution of the CMSSM parameter space (about \fpPercentage\%)
should be 
within the reach of very near future direct detection experiments, and should
be detectable in the near future by IceCube. 

Voltaire's satirical philosopher Pangloss long held the position that we live
in the ``best of all possible worlds''. 
We find that if naturalness considerations are taken into account, 
then (modulo the usual astrophysical uncertainties) the
prospects for the direct and indirect detection of neutralino dark matter in
the CMSSM are, if not Panglossian, are at least extremely encouraging.

\acknowledgments
DH is supported by the Fermi Research Alliance, LLC under Contract No. DE-AC02-07CH11359 with the US Department of Energy and by NASA grant NNX08AH34G. 
This work has been partially supported by STFC\@.
The computational work has been performed using the Cambridge eScience CAMGRID
computing facility, with the invaluable help of M.~Calleja.


\begin{thebibliography}{99}




\bibitem{susyreview}
For a review of supersymmetry phenomenology, see:  S.~P.~Martin,
  arXiv:hep-ph/9709356.

\bibitem{gut}
  J.~R.~Ellis, S.~Kelley and D.~V.~Nanopoulos,
  Phys.\ Lett.\ B {\bf 260}, 131 (1991).


\bibitem{neutralinodm}
  H.~Goldberg,
  Phys.\ Rev.\ Lett.\  {\bf 50}, 1419 (1983);
  J.~R.~Ellis, J.~S.~Hagelin, D.~V.~Nanopoulos, K.~A.~Olive and M.~Srednicki,
  Nucl.\ Phys.\ B {\bf 238}, 453 (1984).


\bibitem{wmap5}
 E.~Komatsu {\it et al.},
  [arXiv:0803.0547].


\bibitem{Bertone:2004pz}
  G.~Bertone, D.~Hooper and J.~Silk,
  Phys.\ Rept.\  {\bf 405}, 279 (2005)
  [arXiv:hep-ph/0404175].


\bibitem{gamma}
  L.~Bergstrom, P.~Ullio and J.~H.~Buckley,
  Astropart.\ Phys.\  {\bf 9}, 137 (1998)
  [arXiv:astro-ph/9712318];
  L.~Bergstrom, J.~Edsjo and P.~Ullio,
  Phys.\ Rev.\ Lett.\  {\bf 87}, 251301 (2001)
  [arXiv:astro-ph/0105048];
  V.~Berezinsky, A.~Bottino and G.~Mignola,
  Phys.\ Lett.\ B {\bf 325}, 136 (1994)
  [arXiv:hep-ph/9402215].


\bibitem{neutrinos}
 L.~Bergstrom, J.~Edsjo and P.~Gondolo,
  Phys.\ Rev.\  D {\bf 55}, 1765 (1997)
  [arXiv:hep-ph/9607237];
  Phys.\ Rev.\  D {\bf 58}, 103519 (1998)
  [arXiv:hep-ph/9806293];
  F.~Halzen and D.~Hooper,
  Phys.\ Rev.\  D {\bf 73}, 123507 (2006)
  [arXiv:hep-ph/0510048];
V.~D.~Barger, F.~Halzen, D.~Hooper and C.~Kao,
Phys.\ Rev.\ D {\bf 65}, 075022 (2002).

\bibitem{positron}
E.~A.~Baltz and J.~Edsjo,
Phys.\ Rev.\ D {\bf 59} (1999) 023511
[arXiv:astro-ph/9808243];
 S.~Profumo and P.~Ullio,
  JCAP {\bf 0407}, 006 (2004)
  [arXiv:hep-ph/0406018].


\bibitem{antiproton}
L.~Bergstrom, J.~Edsjo and P.~Ullio,
arXiv:astro-ph/9906034;
A.~Bottino, F.~Donato, N.~Fornengo and P.~Salati,
Phys.\ Rev.\ D {\bf 58}, 123503 (1998);
F.~Donato, N.~Fornengo, D.~Maurin, P.~Salati and R.~Taillet,
arXiv:astro-ph/0306207.


\bibitem{antideu}
F.~Donato, N.~Fornengo and P.~Salati,
 Phys.\ Rev.\  D {\bf 62}, 043003 (2000)
[arXiv:hep-ph/9904481];
  H.~Baer and S.~Profumo,
JCAP {\bf 0512}, 008 (2005)
[arXiv:astro-ph/0510722].



\bibitem{syn}
  E.~A.~Baltz and L.~Wai,
  Phys.\ Rev.\  D {\bf 70}, 023512 (2004)
  [arXiv:astro-ph/0403528];
M.~P.~Blasi, A.~V.~Olinto and C.~Tyler,
Astropart.\ Phys.\  {\bf 18}, 649 (2003)  [arXiv:astro-ph/0202049];
  D.~P.~Finkbeiner,
  arXiv:astro-ph/0409027;
  S.~Colafrancesco, S.~Profumo and P.~Ullio,
  Astron.\ Astrophys.\  {\bf 455}, 21 (2006)
  [arXiv:astro-ph/0507575];
  L.~Bergstrom, M.~Fairbairn and L.~Pieri,
  Phys.\ Rev.\  D {\bf 74}, 123515 (2006)
  [arXiv:astro-ph/0607327].
  D.~Hooper, D.~P.~Finkbeiner and G.~Dobler,
  Phys.\ Rev.\  D {\bf 76}, 083012 (2007),
  arXiv:0705.3655 [astro-ph].





\bibitem{ellis}
  J.~R.~Ellis, S.~Heinemeyer, K.~A.~Olive and G.~Weiglein,
  JHEP {\bf 0605}, 005 (2006)
  [arXiv:hep-ph/0602220];
  J.~R.~Ellis, K.~A.~Olive and V.~C.~Spanos,
  Phys.\ Lett.\  B {\bf 624}, 47 (2005)
  [arXiv:hep-ph/0504196];
  J.~R.~Ellis, S.~Heinemeyer, K.~A.~Olive and G.~Weiglein,
  JHEP {\bf 0502}, 013 (2005)
  [arXiv:hep-ph/0411216];
  J.~R.~Ellis, S.~Heinemeyer, K.~A.~Olive, A.~M.~Weber and G.~Weiglein,
  JHEP {\bf 0708}, 083 (2007)
  [arXiv:0706.0652 [hep-ph]];
  J.~R.~Ellis, K.~A.~Olive, Y.~Santoso and V.~C.~Spanos,
  Phys.\ Rev.\ D {\bf 69} (2004) 095004
  [arXiv:hep-ph/0310356].


\bibitem{Profumo:2004at}
  S.~Profumo and C.~E.~Yaguna,
  Phys.\ Rev.\ D {\bf 70} (2004) 095004
  [arXiv:hep-ph/0407036].


\bibitem{Allanach:2005kz}
  B.~C.~Allanach and C.~G.~Lester,
  Phys.\ Rev.\  D {\bf 73}, 015013 (2006)
  [arXiv:hep-ph/0507283].


\bibitem{Roszkowski:2006mi}
  L.~Roszkowski, R.~R.~de Austri and R.~Trotta,
  JHEP {\bf 0704} (2007) 084
  [arXiv:hep-ph/0611173].

\bibitem{weather}
  B.~C.~Allanach, K.~Cranmer, C.~G.~Lester and A.~M.~Weber,
  JHEP {\bf 0708}, 023 (2007)
  [arXiv:0705.0487 [hep-ph]].

\bibitem{deAustri:2006pe}
  R.~R.~de Austri, R.~Trotta and L.~Roszkowski,
  JHEP {\bf 0605} (2006) 002
  [arXiv:hep-ph/0602028].


\bibitem{Roszkowski:2007fd}
  L.~Roszkowski, R.~Ruiz de Austri and R.~Trotta,
  JHEP {\bf 0707}, 075 (2007)
  [arXiv:0705.2012 [hep-ph]].

\bibitem{Roszkowski:2007va}
  L.~Roszkowski, R.~R.~de Austri, J.~Silk and R.~Trotta,
  arXiv:0707.0622 [astro-ph].

\bibitem{BPMZ}
  D.~M.~Pierce, J.~A.~Bagger, K.~T.~Matchev and R.-J.~Zhang, 
  Nucl.\ Phys.\ B {\bf 491} (1997) 3, [arXiv:hep-ph/9606211].

\bibitem{softsusy}
B.C. Allanach, 
Comput. Phys. Commun. {\bf 143} (2002) 305, [arXiv:hep-ph/0104145].

\bibitem{pdg06}
  W.~-M.~Yao et al. [Particle Data Group], {\em J. Phys.} {\bf G33} (2006) 1.
  and 2007 partial update for 2008.



\bibitem{darkSide}
B.~C.~Allanach, C.~G.~Lester and A.~M.~Weber, 
JHEP {\bf 12} (2006) 065
[arXiv:hep-ph/0609295]


\bibitem{allanachnatural}
  B.~C.~Allanach,
  Phys.\ Lett.\  B {\bf 635}, 123 (2006)
  [arXiv:hep-ph/0601089].

\bibitem{finetuning}
  R.~Barbieri and G.~F.~Giudice,
  Nucl.\ Phys.\ B {\bf 306} (1988) 63;
  B.~de Carlos and J.~A.~Casas,
  Phys.\ Lett.\ B {\bf 309} (1993) 320,
  {[arXiv:hep-ph/9303291]};
  R.~Barbieri and A.~Strumia,
  Phys.\ Lett.\ B {\bf 433} (1998) 63,
  [arXiv:hep-ph/9801353];
  C.~Giusti, A.~Romanano and A.~Strumia.
  Nucl.\ Phys.\  B {\bf 550}, 3 (1999)
  [arXiv:hep-ph/9811386];
  L.~E.~Ibanez and G.~G.~Ross,
  arXiv:hep-ph/0702046.

\bibitem{cdf+dzero-mtop-07}
  [CDF Collaboration and D0 Collaboration],
  arXiv:0803.1683 [hep-ex].

\bibitem{btaunu}
  available at
  http://www.slac.stanford.edu/xorg/hfag/rare/leppho07/radll/index.html

\bibitem{gamb2}
  F.~Mahmoudi,
  JHEP {\bf 12} (2007) 026
  [arXiv:0710.3791].

\bibitem{delms_sm}
  UTfit Collaboration,
  JHEP (2006)
  [hep-ph/0606167].

\bibitem{gm2SM}
K.~Hagiwara, A.D.~Martin,  D.~Nomura, and T.~Teubner, 
(2006) [arXiv:hep-ph/0611102].

\bibitem{mw}
Tevatron Electroweak Working Group \& The CDF Collaboration,
{\em Winter 2007 Conference Note},
http://fcdfwww.fnal.gov/physics/ewk/2007/wmass/.

\bibitem{sinth}
[The ALEPH, DELPHI, L3 and OPAL Collaborations, the LEP Electroweak Working Group], 
arXiv:0712.0929. 

\bibitem{mw2}
Precision Electroweak Measurements and Constraints on the Standard Model,
The LEP Collaboration,
arXiv:0712.0929.

\bibitem{gamb}
  P.~Gambino, U.~Haisch and M.~Misiak,
  Phys.\ Rev.\ Lett.\  {\bf 94} (2005) 061803
  [arXiv:hep-ph/0410155].

\bibitem{slha}
P. Skands {\em et al}, 
JHEP {\bf 0407} (2004)
036, [arXiv:hep-ph/0311123].

\bibitem{micromegas}
G. Bélanger, F. Boudjema, A. Pukhov, A. Semenov, 
arXiv:0803.2360 [hep-ph];
G. Bélanger, F. Boudjema, A. Pukhov, A. Semenov,
Comput.Phys.Commun.176:367-382,2007 hep-ph/0607059; 
 G. B\'{e}langer, F. Boudjema, A. Pukhov and A. Semenov,
 Comput. Phys. Commun. {\bf 174} (2006) 577 [arXiv:hep-ph/0405253];
G. B\'{e}langer, F. Boudjema, A. Pukhov and A. Semenov,
 Comput. Phys. Commun. {\bf 149} (2002) 103 [arXiv:hep-ph/0112278].
\bibitem{gm2exp}
 G.W.~Bennett {\em et al.} [Muon g-2 collaboration],
Phys. Rev. D {\bf73} (2006) 072003 [arXiv:hep-ex/0602035].

\bibitem{superiso}
 F.~Mahmoudi
   Computer Physics Communications (2008),
   [arXiv:0710.2067].

\bibitem{Heinemeyer:2006px}
  The code is forthcoming in a publication by A.~M.~Weber et al.;
  S.~Heinemeyer, W.~Hollik, D.~St\"ockinger, A.~M.~Weber and G.~Weiglein,
JHEP {\bf 08} (2006) 052,
  [arXiv:hep-ph/0604147].

\bibitem{Buras:2002vd}
  A.~J.~Buras, P.~H.~Chankowski, J.~Rosiek and L.~Slawianowska,
  Nucl.\ Phys.\  B {\bf 659}, 3 (2003)
  [arXiv:hep-ph/0210145].

\bibitem{Isidori:2006pk}
  G.~Isidori and P.~Paradisi,
  Phys.\ Lett.\  B {\bf 639}, 499 (2006)
  [arXiv:hep-ph/0605012].

\bibitem{bank}
  B.~C.~Allanach and C.~G.~Lester, 
  arXiv:0705.0486 [hep-ph]

\bibitem{GelmanAndRubin}
A. Gelman and D. Rubin, 
{\em Inference from Iterative Simulation Using Multiple Sequences},
Stat. Sci. {\bf 7} (1992) 457.

\bibitem{feng}
 K.~L.~Chan, U.~Chattopadhyay and P.~Nath,
 Phys.\ Rev.\  D {\bf 58}, 096004 (1998)
 [arXiv:hep-ph/9710473];
  J.~L.~Feng, K.~T.~Matchev and T.~Moroi,
  Phys.\ Rev.\ Lett.\  {\bf 84}, 2322 (2000)
  [arXiv:hep-ph/9908309];
  J.~L.~Feng, K.~T.~Matchev and T.~Moroi,
  Phys.\ Rev.\  D {\bf 61}, 075005 (2000)
  [arXiv:hep-ph/9909334];
  J.~L.~Feng, K.~T.~Matchev and T.~Moroi,
  arXiv:hep-ph/0003138;
D.~Feldman, Z.~Liu and P.~Nath,
 Phys.\ Lett.\  B {\bf 662}, 190 (2008)
 [arXiv:0711.4591 [hep-ph]].

\bibitem{cdms}
  Z.~Ahmed {\it et al.}  [CDMS Collaboration],
  arXiv:0802.3530 [astro-ph].

\bibitem{xenon}
  J.~Angle {\it et al.}  [XENON Collaboration],
  arXiv:0706.0039 [astro-ph].


\bibitem{zeplin}
  G.~J.~Alner {\it et al.},
  Astropart.\ Phys.\  {\bf 28}, 287 (2007)
  [arXiv:astro-ph/0701858];
  G.~J.~Alner {\it et al.}  [UK Dark Matter Collaboration],
  Astropart.\ Phys.\  {\bf 23}, 444 (2005).

\bibitem{edelweiss}
  V.~Sanglard {\it et al.}  [The EDELWEISS Collaboration],
  Phys.\ Rev.\ D {\bf 71}, 122002 (2005)
  [arXiv:astro-ph/0503265].


\bibitem{cresst}
  G.~Angloher {\it et al.},
  Astropart.\ Phys.\  {\bf 23}, 325 (2005)
  [arXiv:astro-ph/0408006].



\bibitem{warp}
  P.~Benetti {\it et al.},
  arXiv:astro-ph/0701286;
  R.~Brunetti {\it et al.},
  New Astron.\ Rev.\  {\bf 49}, 265 (2005)
  [arXiv:astro-ph/0405342].

\bibitem{kims}  
  H.~S.~Lee. {\it et al.}  [KIMS Collaboration],
  Phys.\ Rev.\ Lett.\  {\bf 99}, 091301 (2007)
  [arXiv:0704.0423 [astro-ph]].


\bibitem{coupp}
  W.~J.~Bolte {\it et al.},
  J.\ Phys.\ Conf.\ Ser.\  {\bf 39}, 126 (2006).

\bibitem{jungmanreview}
  G.~Jungman, M.~Kamionkowski and K.~Griest,
  Phys.\ Rept.\  {\bf 267}, 195 (1996)
  [arXiv:hep-ph/9506380].

\bibitem{ascatter}
G.~B.~Gelmini, P.~Gondolo and E.~Roulet,
Nucl.\ Phys.\ B {\bf 351}, 623 (1991);
M.~Srednicki and R.~Watkins,
Phys.\ Lett.\ B {\bf 225}, 140 (1989);
M.~Drees and M.~Nojiri,
Phys.\ Rev.\ D {\bf 48}, 3483 (1993)
[arXiv:hep-ph/9307208];
M.~Drees and M.~M.~Nojiri,
Phys.\ Rev.\ D {\bf 47}, 4226 (1993)
[arXiv:hep-ph/9210272];
J.~R.~Ellis, A.~Ferstl and K.~A.~Olive, 
Phys.~Lett.~B  481, (2000) 304,
[arXiv:hep-ph/0001005].

\bibitem{nuc}
  A.~Bottino, F.~Donato, N.~Fornengo and S.~Scopel,
  Astropart.\ Phys.\  {\bf 18}, 205 (2002)
  [arXiv:hep-ph/0111229];
 Astropart.\ Phys.\  {\bf 13}, 215 (2000)
  [arXiv:hep-ph/9909228];
  J.~R.~Ellis, K.~A.~Olive, Y.~Santoso and V.~C.~Spanos,
  Phys.\ Rev.\ D {\bf 71}, 095007 (2005)
  [arXiv:hep-ph/0502001].

\bibitem{local}
  E.~I.~Gates, G.~Gyuk and M.~S.~Turner,
  Phys.\ Rev.\ D {\bf 53} (1996) 4138
  [arXiv:astro-ph/9508071],
 E.~Gates, G.~Gyuk and M.~S.~Turner,
  arXiv:astro-ph/9704253.


\bibitem{fine}
  A.~Helmi, S.~D.~M.~White and V.~Springel,
  Phys.\ Rev.\ D {\bf 66}, 063502 (2002)
  [arXiv:astro-ph/0201289];
  M.~Vogelsberger, S.~D.~M.~White, A.~Helmi and V.~Springel,
  arXiv:0711.1105 [astro-ph].

\bibitem{nuclear}
  A.~E.~Nelson and D.~B.~Kaplan,
  Phys.\ Lett.\ B {\bf 192}, 193 (1987);
D.~B.~Kaplan and A.~Manohar,
  Nucl.\ Phys.\ B {\bf 310}, 527 (1988);
 A.~Bottino, F.~Donato, N.~Fornengo and S.~Scopel,
  Astropart.\ Phys.\  {\bf 18}, 205 (2002)
  [arXiv:hep-ph/0111229];
  J.~R.~Ellis, K.~A.~Olive, Y.~Santoso and V.~C.~Spanos,
  Phys.\ Rev.\ D {\bf 71}, 095007 (2005)
  [arXiv:hep-ph/0502001].



\bibitem{capture}
A.~Gould, Astrophys.\ J.\ {\bf 388}, 338 (1991).




\bibitem{superk}
  S.~Desai {\it et al.}  [Super-Kamiokande Collaboration],
  Phys.\ Rev.\  D {\bf 70}, 083523 (2004)
  [Erratum-ibid.\  D {\bf 70}, 109901 (2004)]
  [arXiv:hep-ex/0404025].

\bibitem{icecube}
T.~DeYoung  [IceCube Collaboration],
  Int.\ J.\ Mod.\ Phys.\ A {\bf 20}, 3160 (2005);
J.~Ahrens {\it et al.}  [The IceCube Collaboration],
  Nucl.\ Phys.\ Proc.\ Suppl.\  {\bf 118}, 388 (2003)
  [arXiv:astro-ph/0209556].

\bibitem{amanda}
M.~Ackermann {\it et al.}  [AMANDA Collaboration],
  arXiv:astro-ph/0508518; See also the new preliminary results described in the talk by T.~De Young at {\it Neutrino 08, the XXIII International Conference on Neutrino Physics and Astrophysics}, Christchurch, New Zealand (2008).  

\bibitem{baksan}
M.~M.~Boliev {\it et al.}, 
Proc.\ of the Intl. Workshop on Aspects of Dark Matter in Astrophysics and Particle Physics, Heidelberg, Germany, 1996. Edited by H.~V.~Klapdor-Kleingrothaus, Y.~Ramachers. Singapore, World Scientific, 1997.


\bibitem{macro}
M.~Ambrosio {\it et al.}  [MACRO Collaboration],
  Phys.\ Rev.\ D {\bf 60}, 082002 (1999)
  [arXiv:hep-ex/9812020].









\bibitem{gc}
  A.~Cesarini, F.~Fucito, A.~Lionetto, A.~Morselli and P.~Ullio,
  Astropart.\ Phys.\  {\bf 21}, 267 (2004)
  [arXiv:astro-ph/0305075];
  P.~Ullio, L.~Bergstrom, J.~Edsjo and C.~G.~Lacey,
  Phys.\ Rev.\ D {\bf 66}, 123502 (2002)
  [arXiv:astro-ph/0207125];
  G.~Zaharijas and D.~Hooper,
  Phys.\ Rev.\  D {\bf 73}, 103501 (2006)
  [arXiv:astro-ph/0603540].

\bibitem{dwarf}
  N.~W.~Evans, F.~Ferrer and S.~Sarkar,
  Phys.\ Rev.\  D {\bf 69}, 123501 (2004)
  [arXiv:astro-ph/0311145];
  L.~Bergstrom and D.~Hooper,
  Phys.\ Rev.\  D {\bf 73}, 063510 (2006)
  [arXiv:hep-ph/0512317];
  L.~E.~Strigari, S.~M.~Koushiappas, J.~S.~Bullock, M.~Kaplinghat, J.~D.~Simon, M.~Geha and B.~Willman,
  arXiv:0709.1510 [astro-ph].





\bibitem{dodelson}
  S.~Dodelson, D.~Hooper and P.~D.~Serpico,
  arXiv:0711.4621 [astro-ph].

\bibitem{ac}
  F.~Prada, A.~Klypin, J.~Flix, M.~Martinez and E.~Simonneau,
  arXiv:astro-ph/0401512;
  G.~Bertone and D.~Merritt,
  Mod.\ Phys.\ Lett.\ A {\bf 20}, 1021 (2005)
  [arXiv:astro-ph/0504422].

\bibitem{spike}
  P.~Gondolo and J.~Silk,
  Phys.\ Rev.\ Lett.\  {\bf 83}, 1719 (1999)
  [arXiv:astro-ph/9906391];
  P.~Ullio, H.~Zhao and M.~Kamionkowski,
  Phys.\ Rev.\ D {\bf 64}, 043504 (2001)
  [arXiv:astro-ph/0101481];
  G.~Bertone, G.~Sigl and J.~Silk,
  Mon.\ Not.\ Roy.\ Astron.\ Soc.\  {\bf 337}, 98 (2002)
  [arXiv:astro-ph/0203488].






\bibitem{glast}
  N.~Gehrels and P.~Michelson,
  Astropart.\ Phys.\  {\bf 11}, 277 (1999);
  S.~Peirani, R.~Mohayaee and J.~A.~de Freitas Pacheco,
  Phys.\ Rev.\ D {\bf 70}, 043503 (2004)
  [arXiv:astro-ph/0401378].

\bibitem{nfw}
 J.~F.~Navarro, C.~S.~Frenk and S.~D.~M.~White,
  Astrophys.\ J.\  {\bf 462}, 563 (1996)
  [arXiv:astro-ph/9508025];
  J.~F.~Navarro, C.~S.~Frenk and S.~D.~M.~White,
  Astrophys.\ J.\  {\bf 490}, 493 (1997).


\bibitem{act}
F.~Aharonian {\it et al.}  [The HESS Collaboration],
 Astron.\ Astrophys.\  {\bf 425}, L13 (2004)
  [arXiv:astro-ph/0408145];
  J.~Albert {\it et al.}  [MAGIC Collaboration],
  Astrophys.\ J.\  {\bf 638}, L101 (2006)
  [arXiv:astro-ph/0512469].










\bibitem{diffusion}
For a recent review, see:  A.~W.~Strong, I.~V.~Moskalenko and V.~S.~Ptuskin,
  Ann.\ Rev.\ Nucl.\ Part.\ Sci.\  {\bf 57}, 285 (2007)
  [arXiv:astro-ph/0701517].

\bibitem{pamela}
  P.~Picozza {\it et al.},
  Astropart.\ Phys.\  {\bf 27}, 296 (2007)
  [arXiv:astro-ph/0608697].


\bibitem{ams02}
  M.~Sapinski  [AMS Collaboration],
  Acta Phys.\ Polon.\ B {\bf 37}, 1991 (2006);
  C.~Goy  [AMS Collaboration],
  J.\ Phys.\ Conf.\ Ser.\  {\bf 39}, 185 (2006).





\bibitem{silkpos}
  D.~Hooper and J.~Silk,
  Phys.\ Rev.\ D {\bf 71}, 083503 (2005)
  [arXiv:hep-ph/0409104].

















\end{thebibliography}
\end{document}